\documentclass[journal]{IEEEtran}
\usepackage{amsmath,amsthm,amssymb}
\usepackage{mathrsfs,mathtools,mathabx}
\usepackage{bm}
\usepackage{graphicx}
\usepackage{dsfont,enumitem,stackrel}
\usepackage{color}
\usepackage{array}
\usepackage{cases}
\usepackage{pifont}
\usepackage{algorithm,algpseudocode}
\usepackage{algpseudocode}
\usepackage{pgfplots}
\pgfplotsset{compat=newest}
\usepackage{accents}
\usepackage{subfigure}
\usepackage{float}
\usepackage{cite}
\usepackage{xparse}
\usepackage{xcolor}
\usepackage{tikz}
\usetikzlibrary{spy}
\usetikzlibrary{arrows,decorations.pathreplacing,patterns}
\usepackage{hyperref}
\usepackage{booktabs} 
\usepackage[font=scriptsize]{caption}
\usepackage{lipsum}

\definecolor{redjigar}{rgb}{0.9, 0.01, 0.1}
\definecolor{chestnut}{rgb}{0.8, 0.36, 0.36}
\definecolor{airforceblue}{rgb}{0.36, 0.54, 0.66}
\definecolor{cadmiumorange}{rgb}{0.93, 0.53, 0.18}
\definecolor{bleudefrance}{rgb}{0.19, 0.55, 0.91}
\definecolor{carolinablue}{rgb}{0.6, 0.73, 0.89}
\definecolor{blue(ncs)}{rgb}{0.0, 0.53, 0.74}
\definecolor{dodgerblue}{rgb}{0.12, 0.56, 1.0}
\definecolor{cssgreen}{rgb}{0.0, 0.5, 0.0}
\definecolor{cadmiumgreen}{rgb}{0.0, 0.42, 0.24}
\definecolor{cadmiumorange}{rgb}{0.93, 0.53, 0.18}
\definecolor{amaranth}{rgb}{0.9, 0.17, 0.31}
\definecolor{bluegray}{rgb}{0.4, 0.6, 0.8}
\definecolor{cadmiumgreen}{rgb}{0.0, 0.42, 0.24}
\definecolor{amethyst}{rgb}{0.6, 0.4, 0.8}
\definecolor{antiquebrass}{rgb}{0.8, 0.58, 0.46}

\graphicspath{{./Figures/}} 

\newtheorem{prop}{Proposition}

\newtheorem{remark}{Remark}	
\newtheorem{lem}{Lemma}	
\newtheorem{cor}{Corollary}
\theoremstyle{definition}

\newcommand{\Comp}{ChannelComp~}

\begin{document}
%
\bstctlcite{IEEEexample:BSTcontrol}

\title{ ChannelComp: A General Method for  \\ Computation by  Communications}

\author{Saeed Razavikia$^\dagger$,  José Mairton Barros Da Silva Júnior$^{*}$,  Carlo Fischione$^\dagger$
 	\thanks{S. Razavikia and C. Fischione  are with  the School of Electrical Engineering and Computer Science KTH Royal Institute of Technology, Stockholm, Sweden (e-mail: sraz@kth.se, carlofi@kth.se). C. Fischione is also with Digital Futures of KTH. }
 	\thanks{José Mairton B. da Silva Jr. is with the Department of Information Technology, Uppsala University, Sweden (email: mairton.barros@it.uu.se).}
        \thanks{S. Razavikia was supported by the Wallenberg AI, Autonomous Systems and Software Program (WASP).}
 	\thanks{Jose Mairton B. da Silva Jr. was jointly supported by the European Union’s Horizon Europe research and innovation program under the Marie Skłodowska-Curie project FLASH, with grant agreement No 101067652; the Ericsson Research Foundation, and the Hans Werthén Foundation.}
  \thanks{The EU FLASH project, the Digital Futures project DEMOCRITUS, and the Swedish Research Council Project MALEN partially supported this work.}
    \thanks{A preliminary version of this work was presented in part at the IEEE International Conference on Communications, Rome, Italy, May 2023, which appears in this manuscript as reference~\cite{razavikia2023computing}.}
   
\IEEEauthorblockA{$^\dagger$School of Electrical Engineering and Computer Science, KTH Royal Institute of Technology, Stockholm, Sweden\\
$^*$Department of Information Technology, Uppsala University, Sweden\\
Email: \{sraz, carlofi\}@kth.se, mairton.barros@it.uu.se }

}%

\onecolumn
\noindent \LARGE \textbf{IEEE Copyright Notice}

\noindent \normalsize © 2023 IEEE. Personal use of this material is permitted. Permission from IEEE must be obtained for all other uses, in any current or future media, including reprinting/republishing this material for advertising or promotional purposes, creating new collective works, for resale or redistribution to servers or lists, or reuse of any copyrighted component of this work in other works.

\thispagestyle{empty}
\clearpage
\setcounter{page}{1}
\twocolumn
\maketitle

\pagestyle{headings}
\markboth{IEEE TRANSACTION ON COMMUNICATION. PREPRINT VERSION. OCTOBER 2023}
{Shell \MakeLowercase{\textit{et al.}}: A Sample Article Using IEEEtran.cls for IEEE Journals}

\maketitle

\begin{abstract}	
Over-the-air computation (AirComp) is a well-known technique by which several wireless devices transmit by analog amplitude modulation to achieve a sum of their transmit signals at a common receiver. The underlying physical principle is the superposition property of the radio waves. Since such superposition is analog and in amplitude, it is natural that \mbox{AirComp} uses analog amplitude modulations. Unfortunately, this is impractical because most wireless devices today use digital modulations. It would be highly desirable to use digital communications because of their numerous benefits, such as error correction, synchronization, acquisition of channel state information, and widespread use. However, when we use digital modulations for AirComp, a general belief is that the superposition property of the radio waves returns a meaningless overlapping of the digital signals. In this paper, we break through such beliefs and propose an entirely new digital channel computing method named ChannelComp, which can use digital as well as analog modulations. 
We propose a feasibility optimization problem that ascertains the optimal modulation for computing arbitrary functions over-the-air. Additionally, we propose pre-coders to adapt existing digital modulation schemes for computing the function over the multiple access channel. 
The simulation results verify the superior performance of ChannelComp compared to AirComp, particularly for the product functions, with more than $10$ dB improvement of the computation error.

\end{abstract}

\section{Introduction}

To realize the ubiquitous connectivity from the Internet of Things (IoT), the generations of wireless communications have been accompanied by a paradigm shift from human-type communications towards machine-type communications. On the one hand, the number of IoT devices is predicted to reach $75$ billion by $2025$~\cite{Zhu2021MassiveIoT}, much higher than that of mobile phone users. On the other hand, the various IoT applications based on machine learning (ML) are set to emerge in $6$G \cite{tataria20216g}, and these require the collection, transmission, and calculation of enormous amounts of data from many devices. Consequently, extensive connectivity needs to scale up radio and computation resources, which means swamping the capacity of the current systems. To improve the support of emerging compute-intensive ML applications, e.g., virtual reality, edge computing, federated edge learning, the over-the-air computation (AirComp) method is a promising concept to simultaneously collect and compute data at the edge network~\cite{zhu2019broadband,goldenbaum2013harnessing,nazer2007computation,hellstrom2022wireless}. 

AirComp leverages the waveform superposition property of the multi-access channel (MAC) to realize the aggregation of data simultaneously transmitted by devices, allowing each device to access all radio resources, unlike the standard transmit-then-compute scheme. Moreover, AirComp integrates communication and computation steps, providing ultra-fast wireless data aggregation in IoT networks with high-spectrum efficiency. AirComp reduces the required energy of each device for transmission while it can bring a high rate of communication of that device by harnessing interference to help functional computation. Besides preserving the privacy and security of data, the coverage area can also be enlarged since more devices can transmit simultaneously. 
\begin{table}[!t]
\centering
\caption{Reference list of commonly used variables in this survey. Ordered by case and alphabetically.}
\begin{tabular}{|c| c| }
\toprule
\textbf{Variable} & \textbf{Definition} \\
\midrule
$\mathcal{D}_{f}$ & Domain set of function $f$ \\ 
 $\mathcal{R}_f$ &  Range set of function $f$ \\
 $\mathcal{R}_s$ &   Set of all possible constellation points \\
  $\mathscr{T}(\cdot)$ & Tabular mapping to compute function $f$ \\
$\mathscr{E}_k(\cdot)$ &  Encoder at node $k$ \\
$f$ &  Desired function \\
$f^{(i)}$ & Value of output  $i$ of desired function $f$ \\
 $h_k$ & Channel coefficient between node $k$ and the center point  \\
 $K$ & Number of nodes in the network \\
 $q$ & Number of quantization level \\
 $\vec{s}_i$ & The induced value over-the-air corresponding to $f^{(i)}$ \\  
 $\vec{x}_k$ & Modulated signal of node $k$\\
 $\vec{y}$ & The received value at the center point \\
 $\vec{z}$ & Additive white Gaussian noise \\
\bottomrule
\end{tabular}
\label{Tab:List}
\end{table}
\begin{table*}[t]
\centering
\caption{Digital computation over-the-air methods for a star network with K nodes.}
\begin{tabular}{|c|c|c|c| }
\toprule
\textbf{Methods} & \textbf{Bandwidth/Computation} & \textbf{Modulation scheme} & \textbf{Function} \\
\midrule
ChannelComp (proposed)   & 1  & Any digital modulation & Any  \\ 
AirComp\cite{goldenbaum2013harnessing}   & $1$  & Analog amplitude modulation & Nomographic \\ 
OFDMA & $K$ & Any digital modulation & Any   \\
 OBDA \cite{zhu2020one} & $1$  & BPSK and QPSK & Sign \\ 
 OBDA-FSK \cite{csahin2021distributed} & $2$ & FSK  & Sign  \\ 
 OAC-Balanced \cite{sahin2022over} & $\approx\log{q}$  & PAM  &  Mean  \\ 
\bottomrule
\end{tabular}
\label{Tab:Methods}
\end{table*}

\subsection{Literature review}

Two different AirComp approaches exist in the literature: the uncoded analog aggregation~\cite{gastpar2008uncoded,nazer2007computation} and coded digital AirComp~\cite{soundararajan2012communicating,wagner2008rate}. The coded approaches use the linearity of nested lattice codes to encode the input data for obtaining computation over Gaussian channels~\cite{nazer2011compute,goldenbaum2014nomographic}.  The analog AirComp has been widely studied and progressed from different points of view, such as information theory~\cite{nazer2007computation}, signal processing \cite{goldenbaum2013harnessing}, transceiver design~\cite{chen2018over}, channel state information acquisition~\cite{ang2019robust}, or synchronization issues~\cite{goldenbaum2013robust}. Compared with standard schemes, analog technique AirComp can dramatically reduce the required communication resources, particularly in distributed learning, where it has also attracted growing attention for federated edge learning systems~\cite{yang2020federated,amiri2020federated}.

Although AirComp is a promising concept for data aggregation in communication systems (e.g., federated edge learning), it entirely depends on analog communication, which is difficult for reliable communications due to channel ramifications~\cite{sahin2022over}. Furthermore, AirComp needs analog hardware systems for utilizing analog modulations, which is a drawback due to the limited number of current wireless devices that support analog modulations. Accordingly,  the use of digital modulation is deemed more advantageous due to its outstanding properties in channel correction, source, and channel coding, and widespread adoption. This, however, is believed to be extremely difficult due to that the overlapping of digitally modulated signals returns, in general, meaningless or incomprehensible signals for function computation~\cite{zhu2019broadband,wang2022over}.

Recently, there have been some attempts to devise digital aggregation methods, e.g., one-bit broadband digital aggregation (OBDA)~\cite{zhu2020one} and the other-based majority vote frequency-shift keying (FSK)~\cite{csahin2021distributed}. Moreover, in \cite{zhao2021broadband}, a phase asynchronous OFDM-based version of OBDA has been proposed by designing joint channel decoding and aggregation decoders tailored for digital AirComp. Further,  AirComp's non-coherent communication solution for single and multi-cell using pulse-position modulation and FSK have been studied in \cite{sahin2021over,csahin2021distributed,hassan2022multi}.
 All the aforementioned  OBDA studies are limited to specific functions (sign or summation function) or specific ML training procedures (signSGD problem \cite{bernstein2018signsgd} for computing majority vote). 
 Note that \cite{sahin2022over} has proposed to utilize the balanced number systems to compute the summation function while incurring higher bandwidth usage than the standard OBDA method due to allocating frequency for transmitting every quantized level. 
Similarly, another encoding-based numeral system is proposed in \cite{tang2022radix}, where the decimal representation of the input bits is mapped to a pulse amplitude modulation (PAM) symbol to achieve a processing gain. The most representative methods of the AirComp literature are summarised in Table~\ref{Tab:Methods}, including their bandwidth usage, computation, and which functions can be computed.

Consequently, unlike function computations in the AirComp method, existing attempts to use digital modulations with AirComp cannot compute larger classes of functions beyond Nomographic functions \cite{goldenbaum2014nomographic,goldenbaum2013reliable} and are unsuitable for general digital modulations beyond the simple BPSK or FSK.  Indeed, the existing digital aggregation methods enforce analog AirComp over digital communication. Such enforcement works in very few, and particular cases of digital modulations where it appears highly inefficient regarding communication resource usage, such as satellite communications \cite{kodheli2020satellite},  unmanned aerial vehicles \cite{zeng2016wireless},  distributed consensus~\cite{cao2019internet}.

In this paper, we propose a new channel computation method termed \textit{ChannelComp} that is fully compatible with existing digital communication systems, such as those currently available on any smartphone or IoT system.  We investigate the broad set of functions that ChannelComp can compute, and we investigate how \Comp makes it possible for digital modulation schemes to perform computations.

\input{fig/Fig_System_model2}

\subsection{Contributions}
How do we design or adapt modulations so that a valid computation is performed over-the-air? We answer this question by proposing the ChannelComp method. This paper establishes the conditions for computing functions over-the-air using digital modulation. For a given function, these conditions lead to feasibility optimization problems whose solutions give us the parameters of the modulation yielding acceptable computation over-the-air. Moreover, we show how to adapt existing digital or analog modulation schemes using pre-coders to compute the desired function over the MAC. Note that the rationale behind ChannelComp draws from the benefits of digital modulations, yet its scope can also be extended to encompass analog modulations.

Specifically, our contributions are as follows:
\begin{itemize}
    \item \textbf{ChannelComp:} we propose and establish a novel method, the ChannelComp method, to calculate any function over the MAC by digital communications.
    \item \textbf{Simple communication architecture:} one of the key benefits of \Comp is its innate compatibility with existing digital systems. By contrast, if we wish to implement AirComp on existing digital systems, it is necessary to enforce analog communication systems to work on top of current digital systems, which is very difficult and impractical~\cite{sahin2022survey,wang2022over}.
    \item  \textbf{Tractable complexity:}
    we investigate the fundamental limits for valid computations over MAC and devise an optimization to find the parameters of the digital modulation, which is an \textit{NP-hard} problem. To address such complexity, we propose a convex relaxation that can be solved using a solver such as CVX~\cite{grant2014cvx}. 
    \item \textbf{All finite-valued functions:}  \Comp can compute any function with a finite cardinality input domain, which is the case of any digital system because of the digital representation of values. As long as it is possible to count the output of the function, \Comp allows us to compute the desired function. For instance, to compute the function that returns the maximum of its input variables, the AirComp method uses the log sum function to approximate the maximum, which returns an approximate value. By contrast, ChannelComp, for a finite input domain of the function,  can calculate exactly the maximum of the given input variables.
    \item \textbf{Any digital modulations:} \Comp is not only compatible with digital modulation such as quadratic phase shift key (QPSK), quadrature amplitude modulation  (QAM) $4,16,64,\ldots$, but also FSK modulation, amplitude-shift keying (ASK), or phase modulation (PM), and other digital modulations. 
    \item \textbf{Low latency:} \Comp provides a wireless aggregation communication system as fast as the AirComp or even faster. This is because  \Comp adapts the parameters of the digital modulation format such that the receiver computes the desired function, leading to a low latency computation over-the-air. 
\end{itemize}

 Besides the mentioned benefits, in the numerical experiments we present in this paper, ChannelComp outperforms AirComp in terms of computation error for various critical functions while consuming the same communication resources. For example, for computing the product function, ChannelComp obtains a $10$ dB performance improvement compared to AirComp in terms of the normalized mean square error without using analog modulations while relying only on currently widespread digital modulations.

\subsection{Organization of the paper}

The rest of the paper is organized as follows: in Section~\ref{sec:model}, we explain the system model and  present the ChannelComp method. Then, we investigate the fundamental limits of the ChannelComp computation over the MAC in Section~\ref{sec:fun}. Next, based on our obtained fundamental limits, we characterize in Section~\ref{sec:Modulation} how to select the digital modulation formats for computing the desired function over the MAC with AWGN and fading effects over the wireless channel. We further describe the receiver architecture for \Comp in Section~\ref{sec:Reciever}. We present the numerical results  and the performance comparison between ChannelComp, AirComp, and the traditional OFDMA in Section~\ref{sec:Num}. Finally, we conclude the paper in Section~\ref{sec:conclusion}.

\subsection{Notation}

Throughout this paper, scalars are denoted by lowercase letters $x$, vectors and matrices by lower $\bm{x}$, and upper-case boldface letters $\bm{X}$, respectively. We use $\vec{x}$ to represent modulated band-pass signals. Operators are represented by calligraphic notations such as $\mathcal{X}$. The transpose and Hermitian of a matrix $\bm{X}$ are represented by $\bm{X}^{\mathsf{T}}$ and $\bm{X}^{\mathsf{H}}$, respectively. We further use $\otimes$ to show the Kronecker product.  
For a vector $\bm{x}$, $\|\bm{x}\|_1$ and $\|\bm{x}\|_{\infty}$ are defined as the element-wise $\ell_1$ and $\ell_{\infty}$ norms, respectively. We define $\|\bm{X}\|_2$ and $\|\bm{X}\|_{\rm F}$  as the spectral and Frobenius of the matrix $\bm{X}$, respectively. Also, $\|\bm{X}\|_{1 \rightarrow 1}$ is defined as ${\max_{j\in[N]} \sum_{i}|x_{i,j}|}$ where $x_{i,j}$ denotes $(i,j)$ element of matrix $\bm{X}$.

We define $\mathcal{R}_f$ as the range of function $f$, and  its cardinality  by $|\mathcal{R}_f|$.  For an integer $N$, $[N]$ corresponds to the set  $\{1,2,\dots, N\}$. The finite field of size $q$ with subset of the integers  $\{0,1,2,\ldots,q-1\}$ is represented by $\mathbb{F}_q\subset \mathbb{Z}$. We use $\bm{X}\succeq \bm{0}$ to show that $\bm{X}$ is a positive semidefinite matrix.  Finally, we define the $ {\rm diag}: \mathbb{C}^N \mapsto \mathbb{C}^{N\times N}$ operator for a vector $\bm{x} \in \mathbb{C}^{N}$ that puts the input vector on the output matrix main diagonal and zero elsewhere.

\section{System Model and Method Formulation}\label{sec:model}

Consider a communication network in which there exists a computation point (CP) server with $K$ nodes. The nodes communicate with the CP through a shared communication channel. The CP aims to compute the desired function $f(x_1,x_2,\ldots,x_K)$ whose input $x_k \in \mathbb{F}_q$ is a value owned by node $k$. In particular, we assume that the nodes transmit $x_k$'s by digital communications. In such digital communication systems, all the nodes transmit their values over the MAC to compute the function $f$ at the CP. 

To perform a digital transmission, the usual procedure is the following: each value $x_k$ is quantized into a  scalar $\tilde{x}_k:=\mathcal{Q}(x_k)$ with $q$ possible values, where $\mathcal{Q}(\cdot)$ is the quantizer and $q$ equals $2$ to the power of the number of quantization bits. Then, the resultant vector is mapped into the digitally modulated signal $\vec{x}_k$  using encoder $\mathscr{E}_k(\cdot)$, i.e., $\vec{x}_k = \mathscr{E}_k(\tilde{x}_k)$. The signal $\vec{x}_k$ is what node $k$ transmits over the communication channel. Since all the nodes transmit simultaneously\footnote{It is assumed that the synchronization among all the nodes and the CP is perfect. However, the existing techniques of AirComp for solving imperfect synchronization, e.g.,  \cite{saeed2022BlindFed,Henrik2023Blindair} can be applied to our system model due to its similarity to the AirComp model.} and over the same frequency or codes, the CP server receives the summation of all $\vec{x}_k$'s, over the MAC during the one-time slot~\cite{razavikia2023computing}, i.e.    
\begin{align}
    \label{eq:Aggnoise}
      \vec{y} = \sum_{k=1}^{K}h_kp_k\vec{x}_k+ \vec{z},
\end{align}
where $\vec{y}$ is physically generated by the superposition nature of electromagnetic waves, $h_k$ denotes the channel coefficient between node $k$ and the CP server, $p_k$ is the transmit power used by node $k$, and  $\vec{z}$ is a receiver noise that the receivers unavoidably experience. Following the usual modeling in digital communications, we model such noise as an additive white Gaussian noise (AWGN) process with zero-mean and variance $\sigma_z^2$, which is circularly symmetric. 
To compute the function $f$, we need to use mapping (Tabular mapping) $\mathscr{T}(\vec{y})$ based on the resultant constellation diagram of signal $\vec{y}$ (see Figure~\ref{fig:ComSystem}). Note that this system model is identical to the AirComp system model over the MAC~\cite{goldenbaum2013harnessing}, except for digital modulation, instead of analog modulation, to communicate and for the trivial tabular mapping. Hence, this system model facilitates low-latency communication for a large number of nodes, as demonstrated by \cite{zhu2019broadband}. Moreover, the ChannelComp system model can perform \textit{general} function computations faster than AirComp, as the measured value obtained via tabular mapping $\mathscr{T}(\cdot)$ corresponds exactly to the computation result, whereas the desired value in AirComp requires post-coders for computation \cite{goldenbaum2014nomographic}.

\input{fig/Fig_QPSK}
Note that each digital value $\vec{x}_k$ is selected from finite $q$ possible outputs of $\mathcal{Q}(\cdot)$, thereby the received signal $\vec{y}$  has finite constellation points in the absence of AWGN. The summation in \eqref{eq:Aggnoise} induces a specific constellation diagram of $\vec{y}$ that depends on the number of nodes $K$ and on which modulations have been used for each $\vec{x}_k$. Note that such a resulting constellation diagram is a deformation of the original constellation diagram of the transmitting nodes.  Figure~\ref{fig:ConsBPSK} shows this deformation for the case of using QPSK for two nodes, i.e., $K =2$. Recall that our goal is to compute the desired function $f$. Then, the mapping $\mathscr{T}(\vec{y})$ must be determined such that its output approaches the value of function $f$ as much as possible. Mathematically, at time $t$, the modulation vectors $\vec{x}_1,\ldots,\vec{x}_K$  with the map  $\mathscr{T}$ can be found from the following optimization 
\begin{align}
	\label{eq:Associate}
	\mathscr{T}^{\ast}, \vec{x}_1^{\ast},\ldots,\vec{x}_K^{\ast}  = \hspace{-6pt}\underset{\mathscr{T},\vec{x}_1,\ldots,\vec{x}_K}{\rm argmin} \hspace{-1pt} \sum_{\substack{\\ x_1,\ldots,x_K \in \mathcal{D}_f }}\hspace{-15pt}\hspace{-4pt}\Big|f(x_1,\ldots,x_K)\hspace{-3pt}-\hspace{-3pt}\mathscr{T}\big(\vec{y}\big)\Big|^2\hspace{-2pt},	
\end{align}
where $\mathcal{D}_f$ is the domain of function $f$. 
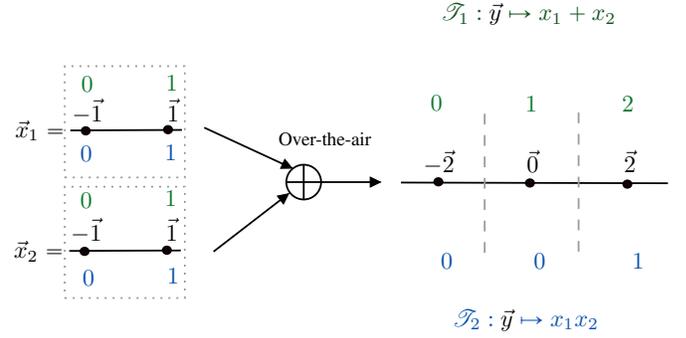
\begin{figure}
    \centering
    
    \scalebox{0.9}{

\tikzset{every picture/.style={line width=0.75pt}} 

\begin{tikzpicture}[x=0.75pt,y=0.75pt,yscale=-0.65,xscale=0.65]

\draw   (259,190) .. controls (259,181.72) and (265.72,175) .. (274,175) .. controls (282.28,175) and (289,181.72) .. (289,190) .. controls (289,198.28) and (282.28,205) .. (274,205) .. controls (265.72,205) and (259,198.28) .. (259,190) -- cycle ; \draw   (259,190) -- (289,190) ; \draw   (274,175) -- (274,205) ;
\draw    (188.25,143) -- (261.28,176.74) ;
\draw [shift={(264,178)}, rotate = 204.8] [fill={rgb, 255:red, 0; green, 0; blue, 0 }  ][line width=0.08]  [draw opacity=0] (8.93,-4.29) -- (0,0) -- (8.93,4.29) -- cycle    ;
\draw    (196.25,250) -- (230.27,223.73) -- (258.63,201.83) ;
\draw [shift={(261,200)}, rotate = 142.32] [fill={rgb, 255:red, 0; green, 0; blue, 0 }  ][line width=0.08]  [draw opacity=0] (8.93,-4.29) -- (0,0) -- (8.93,4.29) -- cycle    ;
\draw    (288.25,190) -- (337.25,190) ;
\draw [shift={(340.25,190)}, rotate = 180] [fill={rgb, 255:red, 0; green, 0; blue, 0 }  ][line width=0.08]  [draw opacity=0] (8.93,-4.29) -- (0,0) -- (8.93,4.29) -- cycle    ;
\draw    (71.83,249.4) -- (168,249) ;
\draw  [fill={rgb, 255:red, 20; green, 1; blue, 1 }  ,fill opacity=1 ] (152.51,248.76) .. controls (152.51,247) and (154.08,245.57) .. (156.03,245.57) .. controls (157.97,245.57) and (159.55,247) .. (159.55,248.76) .. controls (159.55,250.53) and (157.97,251.96) .. (156.03,251.96) .. controls (154.08,251.96) and (152.51,250.53) .. (152.51,248.76) -- cycle ;
\draw  [fill={rgb, 255:red, 20; green, 1; blue, 1 }  ,fill opacity=1 ] (81.57,249.76) .. controls (81.57,248) and (83.14,246.57) .. (85.09,246.57) .. controls (87.03,246.57) and (88.61,248) .. (88.61,249.76) .. controls (88.61,251.53) and (87.03,252.96) .. (85.09,252.96) .. controls (83.14,252.96) and (81.57,251.53) .. (81.57,249.76) -- cycle ;
\draw    (357.83,190.4) -- (588,191) ;
\draw  [fill={rgb, 255:red, 20; green, 1; blue, 1 }  ,fill opacity=1 ] (465.51,190.76) .. controls (465.51,189) and (467.08,187.57) .. (469.03,187.57) .. controls (470.97,187.57) and (472.55,189) .. (472.55,190.76) .. controls (472.55,192.53) and (470.97,193.96) .. (469.03,193.96) .. controls (467.08,193.96) and (465.51,192.53) .. (465.51,190.76) -- cycle ;
\draw  [fill={rgb, 255:red, 20; green, 1; blue, 1 }  ,fill opacity=1 ] (386.57,189.76) .. controls (386.57,188) and (388.14,186.57) .. (390.09,186.57) .. controls (392.03,186.57) and (393.61,188) .. (393.61,189.76) .. controls (393.61,191.53) and (392.03,192.96) .. (390.09,192.96) .. controls (388.14,192.96) and (386.57,191.53) .. (386.57,189.76) -- cycle ;
\draw  [fill={rgb, 255:red, 20; green, 1; blue, 1 }  ,fill opacity=1 ] (549.51,191.76) .. controls (549.51,190) and (551.08,188.57) .. (553.03,188.57) .. controls (554.97,188.57) and (556.55,190) .. (556.55,191.76) .. controls (556.55,193.53) and (554.97,194.96) .. (553.03,194.96) .. controls (551.08,194.96) and (549.51,193.53) .. (549.51,191.76) -- cycle ;
\draw [color={rgb, 255:red, 156; green, 156; blue, 156 }  ,draw opacity=1 ] [dash pattern={on 4.5pt off 4.5pt}]  (430,131) -- (430,252) ;
\draw [color={rgb, 255:red, 156; green, 156; blue, 156 }  ,draw opacity=1 ] [dash pattern={on 4.5pt off 4.5pt}]  (511,130) -- (511,251) ;
\draw    (72.83,145.4) -- (169,145) ;
\draw  [fill={rgb, 255:red, 20; green, 1; blue, 1 }  ,fill opacity=1 ] (153.51,144.76) .. controls (153.51,143) and (155.08,141.57) .. (157.03,141.57) .. controls (158.97,141.57) and (160.55,143) .. (160.55,144.76) .. controls (160.55,146.53) and (158.97,147.96) .. (157.03,147.96) .. controls (155.08,147.96) and (153.51,146.53) .. (153.51,144.76) -- cycle ;
\draw  [fill={rgb, 255:red, 20; green, 1; blue, 1 }  ,fill opacity=1 ] (82.57,145.76) .. controls (82.57,144) and (84.14,142.57) .. (86.09,142.57) .. controls (88.03,142.57) and (89.61,144) .. (89.61,145.76) .. controls (89.61,147.53) and (88.03,148.96) .. (86.09,148.96) .. controls (84.14,148.96) and (82.57,147.53) .. (82.57,145.76) -- cycle ;
\draw  [color={rgb, 255:red, 155; green, 155; blue, 155 }  ,draw opacity=1 ][dash pattern={on 0.84pt off 2.51pt}][line width=0.75]  (68,90) -- (171.5,90) -- (171.5,186) -- (68,186) -- cycle ;
\draw  [color={rgb, 255:red, 155; green, 155; blue, 155 }  ,draw opacity=1 ][dash pattern={on 0.84pt off 2.51pt}][line width=0.75]  (68,194) -- (171.5,194) -- (171.5,290) -- (68,290) -- cycle ;

\draw (250,145) node [anchor=north west][inner sep=0.75pt]  [font=\footnotesize] [align=left] {Over-the-air};
\draw (71.21,219) node [anchor=north west][inner sep=0.75pt]   [align=left] {$-\vec{1}$};
\draw (152.93,219) node [anchor=north west][inner sep=0.75pt]   [align=left] {$\vec{1}$};
\draw (375.21,160) node [anchor=north west][inner sep=0.75pt]   [align=left] {$-\vec{2}$};
\draw (463.93,160) node [anchor=north west][inner sep=0.75pt]   [align=left] {$\vec{0}$};
\draw (547.93,160) node [anchor=north west][inner sep=0.75pt]   [align=left] {$\vec{2}$};
\draw (381,112.4) node [anchor=north west][inner sep=0.75pt]  [color={rgb, 255:red, 8; green, 121; blue, 43 }  ,opacity=1 ]  {$0$};
\draw (463,113.4) node [anchor=north west][inner sep=0.75pt]  [color={rgb, 255:red, 8; green, 121; blue, 43 }  ,opacity=1 ]  {$1$};
\draw (546,113.4) node [anchor=north west][inner sep=0.75pt]  [color={rgb, 255:red, 8; green, 121; blue, 43 }  ,opacity=1 ]  {$2$};
\draw (390,33.4) node [anchor=north west][inner sep=0.75pt]  [color={rgb, 255:red, 15; green, 83; blue, 26 }  ,opacity=1 ]  {$\mathscr{T}_1: {\color{black}\vec{y}} \mapsto x_{1} +x_{2}$};
\draw (22,238.4) node [anchor=north west][inner sep=0.75pt]    {$\Vec{x}_{2} =$};
\draw (72.21,116) node [anchor=north west][inner sep=0.75pt]   [align=left] {$-\vec{1}$};
\draw (153.93,116) node [anchor=north west][inner sep=0.75pt]   [align=left] {$\vec{1}$};
\draw (23,134.4) node [anchor=north west][inner sep=0.75pt]    {$\Vec{x}_{1}=$};
\draw (400,297.4) node [anchor=north west][inner sep=0.75pt]  [color={rgb, 255:red, 8; green, 80; blue, 177 }  ,opacity=1 ]  {$\mathscr{T}_2: {\color{black}\vec{y}} \mapsto x_{1} x_{2}$};
\draw (555,249.4) node [anchor=north west][inner sep=0.75pt]  [color={rgb, 255:red, 8; green, 80; blue, 177 }  ,opacity=1 ]  {$1$};
\draw (470,249.4) node [anchor=north west][inner sep=0.75pt]  [color={rgb, 255:red, 8; green, 80; blue, 177 }  ,opacity=1 ]  {$0$};
\draw (390,249.4) node [anchor=north west][inner sep=0.75pt]  [color={rgb, 255:red, 8; green, 80; blue, 177 }  ,opacity=1 ]  {$0$};
\draw (80,96.4) node [anchor=north west][inner sep=0.75pt]  [color={rgb, 255:red, 8; green, 121; blue, 43 }  ,opacity=1 ]  {$0$};
\draw (153,95.4) node [anchor=north west][inner sep=0.75pt]  [color={rgb, 255:red, 8; green, 121; blue, 43 }  ,opacity=1 ]  {$1$};
\draw (79,196.4) node [anchor=north west][inner sep=0.75pt]  [color={rgb, 255:red, 8; green, 121; blue, 43 }  ,opacity=1 ]  {$0$};
\draw (152,195.4) node [anchor=north west][inner sep=0.75pt]  [color={rgb, 255:red, 8; green, 121; blue, 43 }  ,opacity=1 ]  {$1$};
\draw (79,156.4) node [anchor=north west][inner sep=0.75pt]  [color={rgb, 255:red, 8; green, 80; blue, 177 }  ,opacity=1 ]  {$0$};
\draw (152,155.4) node [anchor=north west][inner sep=0.75pt]  [color={rgb, 255:red, 8; green, 80; blue, 177 }  ,opacity=1 ]  {$1$};
\draw (81,263.4) node [anchor=north west][inner sep=0.75pt]  [color={rgb, 255:red, 8; green, 80; blue, 177 }  ,opacity=1 ]  {$0$};
\draw (154,262.4) node [anchor=north west][inner sep=0.75pt]  [color={rgb, 255:red, 8; green, 80; blue, 177 }  ,opacity=1 ]  {$1$};

\end{tikzpicture}
}

    \caption{The sum and product computation using the BPSK modulation. The correspondence between the constellation points and the corresponding input/output values may differ based on the defined function. The corresponding values for the summation and product functions are represented using green and blue color codes, respectively.   }
    \label{fig:BPSKExample}
\end{figure}
To illustrate optimization problem~\eqref{eq:Associate}, let us consider a simple case with BPSK modulation where the scalar value $x_k$ is one-bit quantized to $\tilde{x}_k \in \{0,1\}$. Then, we have  $\tilde{x}_k = x_k + e_k$, where $e_k$ denotes the quantization error at node $k$. Hence, the resultant modulated symbol is 
 \begin{align}
 \label{eq:BPSK}
 \vec{x}_k = \mathscr{E}_k(\tilde{x}_k) = \begin{cases} 
 				A_c, \quad {\rm if} ~~\tilde{x}_k = 1,\\
 				-A_c, \quad {\rm if} ~~\tilde{x}_k = 0.
 		  \end{cases}
 \end{align}
 Here,  $A_c$ is the amplitude of the carrier signal\footnote{
 We let $\mathscr{E}(\cdot)$ be any modulation, whether it has an analog or digital carrier or if the input of the data is digital or analog. For the digital carriers, such as PAM and pulse position modulation (PPM), we can encode and decode the digital data for each symbol per time. For the analog data,  the quantization step (see Figure~\ref{fig:ComSystem}) allows us to perform the ChannelComp method for either analog modulation, such as amplitude modulation (AM), PM,  and frequency modulations (FM), or digital modulation, such as ASK, PSK, FSK, and QAM.}.  
 Then, assuming a noiseless MAC, the received signal by the CP experiences a constellation formed by the summation of $\vec{y} = \sum_{k =1}^{K}\mathscr{E}_k({x}_k).$ For computing the summation function $f(x_1,x_2)= x_1 + x_2$, the tabular $\mathscr{T}_1$ is a simple map that assigns values of the constellation diagram of $\vec{y}$  to the  corresponding output of $f$, i.e.,  
\begin{align}
\label{eq:Example1}
	\mathscr{T}_1(\{-\vec{2}\}) =  0, \quad\mathscr{T}_1(\{\vec{0}\}) =  1, \quad\mathscr{T}_1(\{\vec{2}\}) =  2,
\end{align}
see Figure~\ref{fig:BPSKExample}. Similarly, for the product function $f(\bm{x})=x_1x_2$,   we can check that this function can be computed by using the following tabular map $\mathscr{T}_2$,
\vspace{-5pt}
\begin{align}
\label{eq:Example2}
	\mathscr{T}_2(\{-\vec{2}\}) =  0, \quad \mathscr{T}_2(\{\vec{0}\}) =  0, \quad\mathscr{T}_2(\{\vec{2}\}) =  1.
\end{align}
This example is shown in Figure~\ref{fig:BPSKExample} using blue color. 
Therefore, we are able to compute a binary summation and product over the noiseless MAC. As long as input data are finite points, the resultant overlapped signal would read to finite constellation points, which lets the desired function $f$ be computed by designing the proper Tabular mapping $\mathscr{T}$ to map the finite constellation points to the output of function $f$. In this regime, the communication system does not need to be changed, and the only requirement is to add a quantizer to the transmitter and a Tabular mapping to the receiver side.

However, if we increase the order of modulation, e.g., to QAM $16$, the constellation points shaped by nodes cannot be uniquely mapped to either summation or product function. Moreover, even for a low-order modulation such as BPSK, computing the simple function $f(x_1,x_2) = x_1 + 2x_2$ is not possible (without errors). In the following section, we overcome this limitation and propose a method to adapt modulations to compute any function.

From the previous examples, the question is: which functions are computable using the \Comp communication architecture? Therefore, in the next section, we investigate the characteristics of the modulation and desired function $f$.

\section{Preliminary Results for noiseless MACs}\label{sec:fun}
In this section, we introduce the basic ideas of ChannelComp, and we give some preliminary results for noiseless MACs, symmetric functions, and the special case where all the nodes use an identical modulation. Based on these results, in the next section, we  present ChannelComp for general functions, not necessarily symmetric, and for the case where the nodes are allowed to use different modulations.

In the following, we establish the important result that when identical modulation encoders are used, i.e., $\mathscr{E}_k(\cdot):= \mathscr{E}(\cdot)$ for all $k \in [K]$, it restricts the degrees of freedom in Equation~\eqref{eq:Aggnoise}. Consequently, this limits the class of functions that can be computed over the MAC.
Hence, utilizing identical modulation enforces  a class of functions $f$ with specific features. Furthermore, we propose a necessary condition on the function $f$ to compute it by ChannelComp  uniquely by using an appropriate tabular mapping $\mathscr{T}$.

\begin{prop}[Necessary condition]
\label{prop:Necessary}
Let the $K$ multivariate function $f(x_1,x_2,\ldots,x_K)$ with domain $\mathcal{D}_f$, where $x_k \in \mathcal{D}_f$ for $k \in [K]$ be a symmetric function, i.e.,   
\begin{align}
\label{eq:permuation}
    f(x_1,\ldots,x_K) = f(\pi(x_1),\ldots,\pi(x_K)),
\end{align}
for all possible permutations by $\pi: \{1,\ldots,K\}\mapsto \{1,\ldots,K\}$.  Let each node use the identical modulation  $\mathscr{E}$. Then, function  $f$ can be computed by the constellation diagram of $\sum_{k=1}^{K}\mathscr{E}({x}_k)$.    
\end{prop}
\begin{proof}
    The proof is by contradiction. We assume that $K=2$, and that $f(x_1,x_2)$ is an asymmetric function, i.e., $f(a,b)\neq f(b,a)$ where $a, b \in \mathcal{D}_f$. Then, for a case where $x_1 =a$ and $x_2 = b$, we have $\vec{a}$ and $\vec{b}$ as modulated signal  and $\vec{a}+\vec{b}$ would be received by the CP, respectively. For the reverse scenario, i.e., $x_1 = b$ and $x_2=a$ the CP also observes $\vec{a}+\vec{b}$. Therefore, we have the same constellation point for different values of the asymmetric function $f$, and it is impossible to assign the same vector $\vec{a}+\vec{b}$ to the two different values of $f(a,b)$ and $f(b,a)$. 
\end{proof}
\begin{cor}\label{cor:Binary}
The condition in Proposition \ref{prop:Necessary} becomes  sufficient for function $f$ with binary domain, i.e., $x_K\in \{0,1\}$. Hence, any symmetric  function $f$ with binary domains can be computed perfectly via one-bit communication.
\end{cor}
\begin{remark}
    In Proposition \ref{prop:Necessary}, we do not make assumptions on the modulation and the domain. Therefore, the proposition is valid for arbitrary modulation with different bits for each node as long as the modulations are identical.    
\end{remark}

To see why the condition in Proposition \ref{prop:Necessary} is not sufficient in general, one can check a simple product function $f(x_1,x_2) = x_1x_2$, where the nodes use the PAM modulation with two bits. Then, some of the function's outputs $f$ overlap their constellation points. Table~\ref{Tab:conterEx} shows this case in detail. In Table~\ref{Tab:conterEx},  $\mathscr{E}(\cdot)$  is the PAM modulation,  and $x_1x_2$ is a symmetric function concerning $x_1$ and $x_2$. This function cannot be computed because a conflict occurs on points $\vec{2}, \vec{3}$ and $\vec{4}$. The constellation points $\vec{2}, \vec{3}$ and $\vec{4}$ need to be assigned simultaneously to the three different values $\{0,1\},\{0,2\}$  and $\{3,4\}$, respectively.

To give more insights, let us consider the summation  $\sum_{k=1}^{K}\mathscr{E}({x}_k)$ as a function that maps the $q^K$ points in the domain of a function to a lower number of constellation points over-the-air. Then, we establish in the following Proposition~\ref{pro:Rang} the lower and upper bound on the cardinality of $\mathcal{R}_s$ when using identical modulation.

\begin{table}[!t]
\centering
\begin{tabular}{l l| l| l}
\toprule
$x_1$ & $x_2$ & $x_1x_2$ & $\vec{x}_1 + \vec{x}_2$\\
\midrule
$0$ & $0$ & $0$ & $\vec{0}$ \\
$0$ & $1$ & $0$ & $\vec{1}$ \\
$1$ & $1$ & ${\color{redjigar}1}$ & ${\color{redjigar}\vec{2}}$ \\
$0$ & $2$ & ${\color{redjigar}0}$ & ${\color{redjigar}\vec{2}}$ \\
$0$ & $3$ & ${\color{redjigar}0}$ & ${\color{redjigar}\vec{3}}$ \\
$1$ & $2$ & ${\color{redjigar}2}$ & ${\color{redjigar}\vec{3}}$ \\
$1$ & $3$ & ${\color{redjigar}3}$ & ${\color{redjigar}\vec{4}}$ \\
$2$ & $2$ & ${\color{redjigar}4}$ & ${\color{redjigar}\vec{4}}$ \\
\bottomrule
\end{tabular}

\caption{A counterexample showing that the symmetric property is insufficient for a correct computation with identical modulations.}
\label{Tab:conterEx}
\end{table}

\begin{prop}\label{pro:Rang}
Let $\mathcal{R}_s$ be the range set of summation  $\sum_{k=1}^{K}\vec{x}_k$ and $\mathcal{R}_f$ be the range of desired function $f$.
Assume that we have $K$ different modulation points $\vec{x}_k, k=1,\ldots, K$, where each one consists of $q$ distinct points in 2D space. Then, the aggregation of the modulated signals, i.e., $\sum_{k=1}^{K}\mathscr{E}({x}_k)$ cannot impose lower than $(q-1)K+1$ number of distinct constellation points and greater than $ \binom{K+q-1}{q-1}$. Therefore, the following lower and upper bounds for the range of summation $\mathcal{R}_s$ holds
 \begin{align}
 (q-1)K+1 \leq |\mathcal{R}_s|\leq \binom{K+q-1}{q-1}. 
 \end{align}   
\end{prop}
\begin{proof}
See appendix \ref{proof:PropRange}. 
\end{proof}
The upper and lower bounds in Proposition \ref{pro:Rang} are the limitations of the range of the symmetric functions. As a result, if the range of the desired function $f$ is greater than the number of the constellation points, i.e., $|\mathcal{R}_f| \geq | \mathcal{R}_s|$,  it is impossible to cover the overall range of function $f$.

Based on the above results,  the use of identical modulation among all the nodes restricts us to compute only the symmetric functions. To go beyond symmetric functions, we need to change the modulations to allow us to compute any class of functions. In the following section, we propose a method to adapt or design the modulations under different scenarios to obtain a reliable function computation over the MAC.

\section{Modulation Selection for Noisy MACs}\label{sec:Modulation}

 We start this section by introducing a general notation to model that each node is allowed to use any modulation, not necessarily identical among the nodes. Moreover, this section assumes that the channel is subject to fading and that the receiver experiences noise. We propose a feasibility check that allows us to design which modulation is needed at each node for a correct computation, which we term as \textit{G-ChannelComp} solution algorithm, where G stands for "general". Then, we introduce how to adapt existing modulations' power and phases, resulting in a correct computation, which we term as \textit{E-ChannelComp} solution algorithm, where E stands for "equal". We conclude the section by giving the general architecture of ChannelComp.
\begin{table}[!t]
\centering
\begin{tabular}{l l| l| l}
\toprule
$x_1$ & $x_2$ & $x_1 + 1$ & $\vec{x}_1 + \vec{x}_2$\\
\midrule
0 & 0 & 1 & $\vec{0}$ \\
0 & 1 & {\color{redjigar}1} & ${\color{redjigar}\vec{1}}$ \\

1 & 0 & {\color{redjigar}2} & ${\color{redjigar}\vec{1}}$ \\
1 & 1 & 2 & $\vec{2}$ \\
\bottomrule
\end{tabular}
\caption{A counter-example to demonstrate that $|R_s|\geq| \mathcal{R}_f|$  is not enough to compute function $f$.}
\label{Tab:NesRebg}
\end{table}

Following the power control universally adopted in the AirComp literature~\cite{cao2019optimal}, i.e., selecting the transmit power as the inverse of the channel as $p_k = h_k^{*}/|h_k|^{2}$,  we rewrite \eqref{eq:Aggnoise} as follows\footnote{The use of this power control is done for analytical simplicity. A more general power selection can also be introduced. The optimization problems we formulate and solve in this section  can be easily extended to include such general power controls. We give an example of this in Section  \ref{sec:AWGN} with optimization problem \eqref{eq:MinimumPwer-nonconvex}.}
\begin{align}
    \label{eq:noisefree}
      \vec{y} = \sum_{k=1}^{K}\vec{x}_k + \vec{z}.
\end{align}
Therefore, the received signal by the  CP, i.e., $\vec{y}$, is equal to $\sum_{k}\vec{x}_k$, which has finite constellation points. Using the complex/real vector representation of each $\bm{x}_k$ for digital/analog modulation, we can represent all the possible constellation points as:
\begin{align}
\label{eq:MatrixFormat}
\vec{\bm{s}}: = \bm{A}\bm{x},    
\end{align}
where $\bm{A} \in \{0,1\}^{{M}\times N}$ ($N:=K\times q$ and $M:=q^K$) is a binary matrix that selects all the  possible cases of nodes to send their bits, and the vector $\bm{x} = [\bm{x}_1,\ldots,\bm{x}_K]^{\mathsf{T}}\in \mathbb{C}^{N \times 1}$ consists of all $K$ blocks of complex vectors, where block $k$ is a modulation of the corresponding node $k$. Also, the vector $\vec{\bm{s}} \in \mathbb{C}^{M\times 1}$ denotes the constellation points over the MAC. 

Now, equipped with the definitions in \eqref{eq:MatrixFormat}, we can establish the conditions to ensure the correctness and viability of computation via wireless transmission and formulate the modulation vector $\bm{x}$ to adhere to these established conditions.

\subsection{Modulation Design with Channel Inversion}

To compute the function $f$ whose input corresponds to the  modulated signal  $\vec{x}_1,\ldots, \vec{x}_K$, we need to make sure that the employed modulations allow such computation with adequate accuracy. As mentioned after Proposition~\ref{pro:Rang}, one necessary condition under which perfect computation can be possible is related to the range of function $R_f$ and the range of the summation over the channel $R_s$. Indeed, the number of constellation points at the receiver CP has to be greater than the range of function $f$,  i.e., $|\mathcal{R}_s| \geq |\mathcal{R}_f|$. To illustrate that this condition is necessary but not sufficient, we can consider function $f$ to be $x_1+1$, for $x_1,x_2\in \{0,1\}$, where the signals are modulated using BPSK. In Table~\ref{Tab:NesRebg}, we illustrated the range points of the function $f$ and  summation over the MAC, i.e., $\sum_{k=1}^K\vec{x}_k$. While we have $|\mathcal{R}_s|=3> 2= |\mathcal{R}_f|$, the function $f$ cannot be computed by this modulation because of the overlapping on point $\vec{1}$.

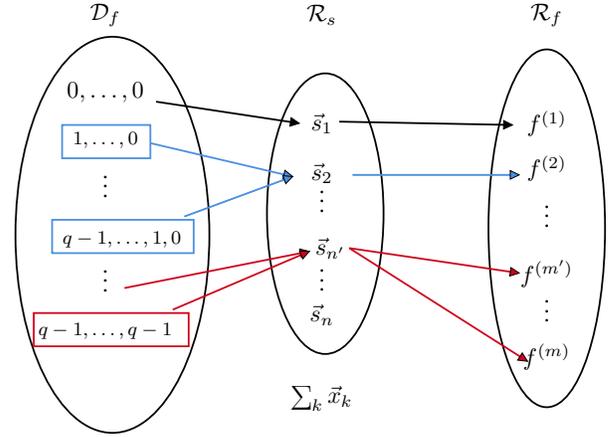
\begin{figure}
    \centering
  \scalebox{0.9}{  

\tikzset{every picture/.style={line width=0.75pt}} 

\begin{tikzpicture}[x=0.7pt,y=0.7pt,yscale=-1,xscale=1]

\draw   (86,152) .. controls (86,86.28) and (112.06,33) .. (144.21,33) .. controls (176.37,33) and (202.43,86.28) .. (202.43,152) .. controls (202.43,217.72) and (176.37,271) .. (144.21,271) .. controls (112.06,271) and (86,217.72) .. (86,152) -- cycle ;
\draw   (370,148.1) .. controls (370,88.73) and (385.67,40.6) .. (405,40.6) .. controls (424.33,40.6) and (440,88.73) .. (440,148.1) .. controls (440,207.47) and (424.33,255.6) .. (405,255.6) .. controls (385.67,255.6) and (370,207.47) .. (370,148.1) -- cycle ;
\draw   (237,139.3) .. controls (237,92.74) and (252.67,55) .. (272,55) .. controls (291.33,55) and (307,92.74) .. (307,139.3) .. controls (307,185.86) and (291.33,223.6) .. (272,223.6) .. controls (252.67,223.6) and (237,185.86) .. (237,139.3) -- cycle ;
\draw    (170.43,72) -- (253.46,84.55) ;
\draw [shift={(256.43,85)}, rotate = 188.6] [fill={rgb, 255:red, 0; green, 0; blue, 0 }  ][line width=0.08]  [draw opacity=0] (5.36,-2.57) -- (0,0) -- (5.36,2.57) -- cycle    ;
\draw [color={rgb, 255:red, 74; green, 144; blue, 226 }  ,draw opacity=1 ]   (167.43,97) -- (249.51,116.31) ;
\draw [shift={(252.43,117)}, rotate = 193.24] [fill={rgb, 255:red, 74; green, 144; blue, 226 }  ,fill opacity=1 ][line width=0.08]  [draw opacity=0] (5.36,-2.57) -- (0,0) -- (5.36,2.57) -- cycle    ;
\draw [color={rgb, 255:red, 74; green, 144; blue, 226 }  ,draw opacity=1 ][fill={rgb, 255:red, 74; green, 144; blue, 226 }  ,fill opacity=1 ]   (187.43,141) -- (249.61,118.04) ;
\draw [shift={(252.43,117)}, rotate = 159.73] [fill={rgb, 255:red, 74; green, 144; blue, 226 }  ,fill opacity=1 ][line width=0.08]  [draw opacity=0] (5.36,-2.57) -- (0,0) -- (5.36,2.57) -- cycle    ;
\draw [color={rgb, 255:red, 74; green, 144; blue, 226 }  ,draw opacity=1 ][fill={rgb, 255:red, 74; green, 144; blue, 226 }  ,fill opacity=1 ]   (288.43,116) -- (385.43,116) ;
\draw [shift={(388.43,116)}, rotate = 180] [fill={rgb, 255:red, 74; green, 144; blue, 226 }  ,fill opacity=1 ][line width=0.08]  [draw opacity=0] (5.36,-2.57) -- (0,0) -- (5.36,2.57) -- cycle    ;
\draw [color={rgb, 255:red, 208; green, 2; blue, 27 }  ,draw opacity=1 ]   (286.43,160) -- (385.46,174.56) ;
\draw [shift={(388.43,175)}, rotate = 188.37] [fill={rgb, 255:red, 208; green, 2; blue, 27 }  ,fill opacity=1 ][line width=0.08]  [draw opacity=0] (5.36,-2.57) -- (0,0) -- (5.36,2.57) -- cycle    ;
\draw [color={rgb, 255:red, 208; green, 2; blue, 27 }  ,draw opacity=1 ]   (286.43,160) -- (390.9,226.39) ;
\draw [shift={(393.43,228)}, rotate = 212.44] [fill={rgb, 255:red, 208; green, 2; blue, 27 }  ,fill opacity=1 ][line width=0.08]  [draw opacity=0] (5.36,-2.57) -- (0,0) -- (5.36,2.57) -- cycle    ;
\draw    (280.43,84) -- (381.43,85.94) ;
\draw [shift={(384.43,86)}, rotate = 181.1] [fill={rgb, 255:red, 0; green, 0; blue, 0 }  ][line width=0.08]  [draw opacity=0] (5.36,-2.57) -- (0,0) -- (5.36,2.57) -- cycle    ;
\draw [color={rgb, 255:red, 208; green, 2; blue, 27 }  ,draw opacity=1 ]   (151.43,184) -- (259.49,162.58) ;
\draw [shift={(262.43,162)}, rotate = 168.79] [fill={rgb, 255:red, 208; green, 2; blue, 27 }  ,fill opacity=1 ][line width=0.08]  [draw opacity=0] (5.36,-2.57) -- (0,0) -- (5.36,2.57) -- cycle    ;
\draw [color={rgb, 255:red, 208; green, 2; blue, 27 }  ,draw opacity=1 ]   (180.43,197) -- (259.67,163.18) ;
\draw [shift={(262.43,162)}, rotate = 156.89] [fill={rgb, 255:red, 208; green, 2; blue, 27 }  ,fill opacity=1 ][line width=0.08]  [draw opacity=0] (5.36,-2.57) -- (0,0) -- (5.36,2.57) -- cycle    ;

\draw (140,66.4) node  {$0,\dotsc ,0$};
\draw (140,118.4) node    {$\vdots $};
\draw (140,175.4) node     {$\vdots $};
\draw (270,85) node    {$\vec{s}_{1}$};
\draw (270,115) node   {$\vec{s}_{2}$};
\draw (275,160) node    {$\vec{s}_{n'}$};
\draw (270,127.4) node     {$\vdots $};
\draw (270,200) node   {$\vec{s}_{n}$};
\draw (270,175) node    {$\vdots $};
\draw (140,20) node    {$\mathcal{D}_{f}$};
\draw (270,20) node    {$\mathcal{R}_{s}$};
\draw (405,20) node   {$\mathcal{R}_{f}$};
\draw (270,250) node     {$\sum _{k}\vec{x}_{k}$};
\draw  [color={rgb, 255:red, 74; green, 144; blue, 226 }  ,draw opacity=1 ]  (114,86) -- (167,86) -- (167,106) -- (114,106) -- cycle  ;
\draw (140,95) node  [font=\footnotesize]  {$1,\dotsc ,0$};
\draw  [color={rgb, 255:red, 74; green, 144; blue, 226 }  ,draw opacity=1 ]  (108,143) -- (193,143) -- (193,163) -- (108,163) -- cycle  ;
\draw (150,155) node   [font=\footnotesize]  {$q-1,\dotsc ,1,0$};
\draw  [color={rgb, 255:red, 208; green, 2; blue, 27 }  ,draw opacity=1 ]  (97,199) -- (190,199) -- (190,219) -- (97,219) -- cycle  ;
\draw (140,210) node  [font=\footnotesize]  {$q-1,\dotsc ,q-1$};
\draw (405,85) node    {$f^{(1)}$};
\draw (405,112) node    {$f^{(2)}$};
\draw (405,225) node     {$f^{(m)}$};
\draw (405,193.4) node     {$\vdots $};
\draw (405,175) node   {$f^{(m')}$};
\draw (405,136.4) node   {$\vdots $};

\end{tikzpicture}
    }
    \caption{This figure gives the main idea of ChannelComp. It reports the domain $\mathcal{D}_f$, the range of the summation $\mathcal{R}_s$ returned by a noise-free channel, and the desired function $\mathcal{R}_f$ on left, middle, and right, respectively. Two different points (blue lines) in the domain of function $f$ create a constellation point $\vec{s}_2$ while the function's output for these two points is the same and equal to $f^{(2)}$. However, for $\vec{s}_{n'}$ the corresponding values of the function are different $f^{(m)} \neq f^{(m')}$. Accordingly, we cannot assign  the point $\vec{s}_{n'}$ to these points (red lines) unless we enforce a splitting of $\vec{s}_{n'}$ by a proper selection of the modulations.}
    \label{fig:range_vector}
\end{figure}

Furthermore, the sum $\sum_{k=1}^K\vec{x}_k$ must cover the entire range of function $f$. In fact, for $i\in [M]$, let $f^{(i)}$ be an output of function $f$ for a certain value of inputs $x_1,x_2\ldots,x_K$, where all $x_k$'s have the same $q$ possible values. Then, if $f^{(i)}$ is different from $f^{(j)}$,  the corresponding constellation point $\vec{s}_i$ must not be the same as $\vec{s}_j$ for $i\neq j$, see Figure~\ref{fig:range_vector}. Hence, to verify if  given modulation signals are suitable for computing the desired function $f$, we pose the following problem:
\begin{subequations}
\label{eq:feasibility}
\begin{align}
\nonumber
 \mathcal{P}_1 = &{\rm find} ~~\bm{x}~~\\
&~{\rm s.t.}~~{\rm if~}
f^{(i)}\neq f^{(j)} \Rightarrow \vec{s}_i \neq \vec{s}_j,~ 
 \forall (i,j) \in [M]^2,\\& ~~~ \|\bm{x}\|_2^2 \leq 1.
\end{align}
\end{subequations}
Recall that $\bm{x} \in \mathbb{C}^{N\times 1}$ is the complex modulation vector, and $[M]^2$ means $[M] \times [M]$. Problem $\mathcal{P}_1$ is a feasibility problem to find a point that satisfies the constraints, which are non-convex and non-smooth. To simplify problem $\mathcal{P}_1$, we replace such constraints by a smooth condition, 
\begin{subequations}
\label{eq:feasibility-convex}
\begin{align}
\nonumber
 \mathcal{P}_2 = &{\rm find} ~~\bm{x} ~~\\
 &{\rm s.t.}~~
|\vec{s}_i-\vec{s}_j|^2  \geq \epsilon |f^{(i)}- f^{(j)}|^2,\forall (i,j) \in  [M]^2, \\ & ~~~ \|\bm{x}\|_2^2 \leq 1,
\end{align}
\end{subequations}
where $\epsilon > 0$ is a positive constant. Note that for any sufficiently small $\epsilon$, the solution to Problem $\mathcal{P}_1$ equals the solution to Problem $\mathcal{P}_2$. 
\begin{remark}
Problem  $\mathcal{P}_2$ not only satisfies the constraints of Problem  $\mathcal{P}_1$, but also designs the transmit constellation points for achieving the acceptable computation error in noisy communications. The reason is that the right side of the constraints is the computation error, in which a higher computation error due to noise  enforces a larger distance of constellation points and enforces more energy. In other words, the distances among constellation points are penalized based on possible computation errors.
\end{remark}
\begin{remark}
    \label{rem:P2}
    The constraints in Problem  $\mathcal{P}_2$ are compatible with digital or analog modulation, where vector $\bm{x}$ must be a complex or real vector, respectively. Also, the obtained modulation can compute a larger class of functions than Nomographic functions. The solution to Problem  $\mathcal{P}_2$ for the special case of the summation function, i.e., $f = \sum_{k=1}^Kx_k$, results in AirComp. Indeed, for the computation of Nomographic functions, pre-coder, and post-coder functions are utilized in the AirComp method~\cite{Golden2013Harnessing}, which can also be incorporated at the beginning and end of our system model illustrated in Figure \ref{fig:ComSystem}.
\end{remark}
To determine how small $\epsilon$ must be such that the solution to Problem $\mathcal{P}_2$ is equal to the solution to Problem $\mathcal{P}_1$, we have the following lemma.  
\begin{lem}
\label{lem:Lagrangian}
Let $\epsilon^{-1}\geq \max_{(i,j)\in [M]^2} |f^{(i)}-f^{(j)}|^2.$ 
Then, Problem $\mathcal{P}_2$ is feasible,  and thus there exists a modulation vector $\bm{x}$ satisfying the constraints.
\end{lem}
\begin{proof}
See Appendix \ref{sec:Prooflem1}. 
\end{proof}
If we rewrite Problem $\mathcal{P}_2$  in terms of variable $\bm{x}$, we have
\begin{subequations}
\label{eq:MinPower-nonconvex}
\begin{align}
\nonumber
 \mathcal{P}_2 = &{\rm find} ~~ \bm{x} \\ 
 & ~~{\rm s.t.}~~~|(\bm{a}_i - \bm{a}_j)^{\mathsf{T}}\bm{x}|^2\geq \epsilon|f^{(i)}- f^{(j)}|^2, \\& ~~\|\bm{x}\|_2^2 \leq 1, 
\end{align}
\end{subequations}
for all $(i,j) \in [M]^2, i \neq j$.  Problem $\mathcal{P}_2$ is now a quadratically constrained quadratic programming (QCQP) problem with non-convex constraints. From the literature~\cite{Sidir2006Physical}, we note that Problem $\mathcal{P}_2$ is \textit{NP-hard}.
Toward overcoming the non-convex constraints, we can use the \textit{lifting trick}~\cite{vandenberghe1996semidefinite}, in which  we recast the constraints \eqref{eq:MinPower-nonconvex} as 

\begin{algorithm}[!t]
\caption{ Approximated  G-ChannelComp (solution to $\mathcal{P}_4$)}\label{alg:DC}
\begin{algorithmic}[1]
	\State \textbf{Input:} Function $f(x_1,\ldots,x_K)$, $\delta$  
	\State Set $N = qK$, $M = q^K$
	\State \textbf{Output:} Modulation vector $\bm{x} \in \mathbb{C}^{N}$
	\Procedure{Dual-primal DC}{$\{f^{(i)}\}_{i=1}^{M} $}
	\For {$t \gets 1,2,\ldots,$}
	\State Compute Eigendecomposition $\bm{X}^{t-1} = \bm{U} \bm{\Lambda}\bm{U}^{\mathsf{H}}$
	\State Set $\bm{G} = \bm{u}_1\bm{u}_1^{\mathsf{H}}$ where $\bm{u}_1$ is the first column of matrix $\bm{U}$
	\State Compute $\bm{Q} = \bm{G} +  \mu \bm{X}^{t-1}$
	\State Obtain $\bm{X}^{t} =  \underset{\bm{X}\in \mathcal{C}}{\rm argmin}~~   \frac{\mu}{2} \|\bm{X}\|_{\rm F}^2+  \langle \bm{X}, \bm{I}_N -  \bm{Q} \rangle$
    \State Compute  $ \theta = \frac{\mu}{2} \|\bm{X}^{t}\|_{\rm F}^2+  \langle \bm{X}^{t}, \bm{I}_N - \bm{Q}\rangle$
	\If{$|\theta| \leq \delta$}
	\State $\bm{X}^* = \bm{X}^t$
	\State \textbf{break}
	\EndIf
	\EndFor
	\EndProcedure
\end{algorithmic}
\end{algorithm}
\begin{subequations}
\label{eq:tracePwer-nonconvex}
\begin{align}
\nonumber
\mathcal{P}_2 &= {\rm find} ~~~ \bm{x}~~ \\
&~~{\rm s.t.}~~
 {\rm trace}(\bm{x}\bm{x}^{\mathsf{H}}\bm{B}_{i,j})  \geq \gamma_{i,j},
 \\ & ~~{\rm trace}(\bm{x}\bm{x}^{\mathsf{H}}) \leq 1,
\end{align}
\end{subequations}
where $ \gamma_{i,j} = \epsilon |f^{(i)}-f^{(j)}|^2$  for all $(i,j) \in [M]^2, i \neq j$ and $\bm{B}_{i,j} = (\bm{a}_i-\bm{a}_j)(\bm{a}_i-\bm{a}_j)^{\mathsf{T}}$. Note that problem~\eqref{eq:tracePwer-nonconvex} is equivalent to problem~\eqref{eq:feasibility-convex}, and thus we denote problem  \eqref{eq:tracePwer-nonconvex} as $\mathcal{P}_2$. Now, if we consider $\bm{x}\bm{x}^{\mathsf{H}}$ as a matrix  $\bm{X}\in \mathbb{C}^{N\times N}$, then problem~\eqref{eq:tracePwer-nonconvex} can be reformulated as 
\begin{subequations}
\label{eq:traceX-nonconvex}
\begin{align}
 \nonumber
\mathcal{P}_2 &= {\rm find} ~~ \bm{X}\\ 
&~~{\rm s.t.}~~{\rm trace}( \bm{X}\bm{B}_{i,j}^{\mathsf{T}})   \geq \gamma_{i,j},~ {\rm trace}(\bm{X}) \leq 1, \\
& ~~ \bm{X} \succeq \bm{0}, \quad {\rm rank}(\bm{X}) = 1,
\end{align}
\end{subequations}
in which the inequality $\bm{X} \succeq \bm{0}$ means that the matrix $\bm{X}$ is symmetric positive semi-definite (PSD). Constraints~\eqref{eq:traceX-nonconvex} are linear and convex with respect to matrix $\bm{X}$, except for the non-convex rank. One possible way to handle such tricky rank constraint is to relax problem~\eqref{eq:traceX-nonconvex} by dropping the rank-one constraint in \eqref{eq:traceX-nonconvex}~\cite{Sidir2006Physical}, which leads to
\begin{subequations}
\label{eq:traceX-convex}
\begin{align}
\nonumber
\mathcal{P}_3 & ={\rm find} ~~\bm{X}~~ \\
&~~{\rm s.t.}~~ {\rm trace}( \bm{X}\bm{B}_{i,j}^{\mathsf{T}})  \geq \gamma_{i,j}, ~{\rm trace}(\bm{X}) \leq  1\\
& ~~~~ \bm{X} \succeq \bm{0}.
\end{align}
\end{subequations}
Problem $\mathcal{P}_3$ is a semi-definite programming problem that can be solved using CVX~\cite{grant2014cvx}. Moreover, we term the solution algorithm to problem $\mathcal{P}_3$ as the G-ChannelComp methods. Afterward, if $\bm{X}^*$, as the solution to Problem $\mathcal{P}_3$, results to be a rank one matrix, $\bm{x}^{*}$,  the optimal modulation vector solution to Problem $\mathcal{P}_2$ can be obtained via Cholesky decomposition of  $\bm{X}^*$~\cite{luo2010semidefinite}. Otherwise, if the solution to  $\mathcal{P}_3$ does not give a rank one matrix, we can recover a sub-optimal solution to Problem $\mathcal{P}_2$ by the Gaussian randomization method~\cite{luo2010semidefinite}.  The Gaussian randomization method has a guaranteed optimality gap~\cite{Miarton2021Full}. The suboptimality of Problem $\mathcal{P}_3$  may occur when the dimension of the optimization variables becomes large, i.e., $N\gg 1$, which means $K\gg 1 $ or $q\gg 1$. To overcome such drawback, the rank-one constraint can be replaced by another equivalent function, which would be successively solved using the primal and dual problems, see~\cite{wang2020wireless}. Particularly, the rank-one constraint can be translated to ${\rm trace}(\bm{X}) - \|\bm{X}\|_2$. Consequently, we can replace the rank-one constraint in \eqref{eq:traceX-nonconvex} with a penalty term as follows
\begin{subequations}
\label{eq:Penalty-convex}
\begin{align}
\nonumber \mathcal{P}_4 &= \underset{\bm{X}}{\rm min}~~{\rm trace}(\bm{X}) - \|\bm{X}\|_2 \\ 
& ~~{\rm s.t.}~~ 
{\rm trace}( \bm{X}\bm{B}_{i,j}^{\mathsf{T}}) \geq \gamma_{i,j},\\ 
&~~~ \bm{X} \succeq \bm{0}, \quad {\rm trace}(\bm{X}) \leq 1.
\end{align}
\end{subequations}
As Problem $\mathcal{P}_4$ remains non-convex, the optimization can be solved using the difference-of-convex (DC) programming~\cite{wang2020wireless}. The DC algorithm for Problem $\mathcal{P}_4$ is presented in Algorithm~\ref{alg:DC}. 

The relaxation in $\mathcal{P}_3$ disregards the rank constraint in $\mathcal{P}_2$ and can occasionally result in a suboptimal solution, as our manuscript acknowledges. However, when the solution to $\mathcal{P}_3$ (denoted as $\hat{\bm{X}}$) is a rank-one matrix, it is considered optimal and not suboptimal with respect to problem  $\mathcal{P}_2$. In fact, $\mathcal{P}_3$ has the potential to provide optimal solutions to $\mathcal{P}_2$ under certain conditions, particularly when $q$ (level of quantization) or $K$ (number of nodes) are small. The reason is that $\mathcal{P}_3$ is a relaxation of $\mathcal{P}_2$, and if a relaxed problem gives an optimal solution that is feasible to the original problem $\mathcal{P}_2$ (that is defined as a subset of $\mathcal{P}_3$), then it is also optimal to problem $\mathcal{P}_2$. Thus, the rank-one $\hat{\bm{X}}$ from  $\mathcal{P}_3$ is the optimal solution to $\mathcal{P}_2$. 
    
To counter the potential suboptimality of $\mathcal{P}_3$, we proposed $\mathcal{P}_4$ aiming to ensure a low-rank solution as much as possible. This approach is designed to verify the feasibility of rank-one constraints accurately and encourage rank-one solutions. As such, if the solution is a rank-one matrix, it can be regarded as an optimal solution satisfying the constraints of $\mathcal{P}_4$. The main advantage of using $\mathcal{P}_4$ is its ability to reach the stationary point solution, albeit with higher iteration and complexity than $\mathcal{P}_3$. However, since $\mathcal{P}_4$ is not a convex problem, the uniqueness of the solution cannot be guaranteed.

\begin{remark}
Optimization Problem $\mathcal{P}_4$ presented in \eqref{eq:Penalty-convex} must be solved only once and offline at the CP. To compute the solution, the CP only needs to know which function $f$ is in use and how many nodes are present. Once the CP solves optimization Problem $\mathcal{P}_4$ and sends the encoder $\mathscr{E}_k(\cdot)$ to the nodes, node $k$ can utilize its modulation vector $\vec{x}_k$ to communicate and compute the intended function $f$ over the MAC. Therefore, no optimization must be solved during the computation over-the-air, and the modulation and encoder blocks are integrated.
\end{remark}

\subsection{Adapting Phase and Power of Existing Modulations}\label{sec:AWGN}

To this point, we have established a method for designing the modulation vector for computation over the MAC. However, altering the nodes' modulation formats may sometimes be difficult. In situations where the modulation formats are fixed, we show that it is possible to alter the power and phase of each modulation vector to reshape the constellation diagram and ensure correct and feasible function computation. In order to do so, we do not assume a power control as in AirComp that enforces channel inversion in \eqref{eq:MatrixFormat}, but we introduce a general power control. 
Consequently,  the received signal at the CP is
\begin{align}
\label{eq:noisy_measure}
    \vec{y}_n = \sum_{k=1}^K h_k p_k\vec{x}_k + \vec{z},
\end{align}
where $p_k\in \mathbb{C}$ determines the power and the phase for node $k$, $h_k$ is the  channel coefficient form node $k$ to CP, and $\vec{z}\in \mathbb{C}$ denotes circularly symmetric AWGN noise with zero mean and variance $\sigma_z^2$.  To alleviate the notation,  let us define vectors $\bm{p} := [p_1,\ldots,p_K]^{\mathsf{T}} \in \mathbb{C}^{K}$ and $\bm{h} := [h_1,\ldots,h_K]^{\mathsf{T}} \in \mathbb{C}^{K}$. Also, we need to define the operator $\mathcal{H}_q:\mathbb{C}^{K}\mapsto \mathbb{C}^{N\times N}$  as follows
\begin{align}
    \mathcal{H}_q(\bm{h}) := {\rm diag}(\bm{h}) \otimes \bm{I}_{q} = {\rm diag}\big(h_1\bm{I}_{q},\ldots,h_K\bm{I}_{q}\big),
\end{align}
where $\otimes$ denotes Kronecker product and $\bm{I}_{q}$ represents the $q\times q$ identity matrix. Then, the constellation points induced by $\bm{x}$ can be represented as 
\begin{align}
    \bm{A}\underbrace{\mathcal{H}(\bm{h}){\rm diag}(\bm{x}) (\bm{I}_K \otimes\mathds{1}_q )}_{:=\bm{C}}\bm{p} = \tilde{\bm{s}},
\end{align}
where $\mathds{1}_q$ stands for a vector of  size $q\times 1$ whose elements are one, and  $\tilde{\bm{s}}= [\tilde{s}_1,\tilde{s}_2,\ldots,\tilde{s}_M]^{\mathsf{T}}\in  \mathbb{C}^{M}$ denotes the constellation points in the presence of fading. Also, $\bm{x}$ is the given modulation whose elements can be either complex-valued (digital modulation) or real-valued (analog modulation). 

 Accordingly, by incorporating the constraints in \eqref{eq:MinPower-nonconvex},  we propose the following optimization problem to find the minimum modulation powers $\bm{p}$ for achieving a correct computation without changing the modulation formats:  
\begin{align}
\nonumber
\mathcal{P}_5 = & \underset{\bm{p}\in \mathbb{C}^K}{\rm min} ~~\|\bm{p}\|_2^2\quad   \\ \label{eq:MinimumPwer-nonconvex}
& ~~{\rm s.t.}~~|(\bm{a}_i - \bm{a}_j)^{\mathsf{T}}\bm{C}\bm{p}|^2\geq \gamma_{i,j},~\forall~(i,j)\in [M]^2.
\end{align}

Problem $\mathcal{P}_5$ is a QCQP problem and difficult to solve. Therefore, by following similar steps from \eqref{eq:tracePwer-nonconvex} to \eqref{eq:traceX-convex}, we can pose the following relaxed version of Problem $\mathcal{P}_5$:
\begin{align}
    \nonumber
\mathcal{P}_6 & = \underset{\bm{P}\in \mathbb{C}^{K \times K}}{\rm min} 
{\rm trace}(\bm{P}) \quad 
\\ \label{eq:traceP-convex}& ~~ {\rm s.t.}~~{\rm trace}( \bm{P}\bm{C}_{i,j}^{\mathsf{T}})  \geq \gamma_{i,j},~ \bm{P} \succeq \bm{0},
\end{align}
where $\bm{C}_{i,j}:= \bm{C}(\bm{a}_i-\bm{a}_j)(\bm{a}_i-\bm{a}_j)^{\mathsf{T}}\bm{C}^{\mathsf{T}}$. Similar to Problem $\mathcal{P}_3$, the solution to Problem $\mathcal{P}_6$, denoted as  $\bm{P}^*$, can give $\bm{p}^*$ as an optimal solution to $\mathcal{P}_5$ by using Cholesky decomposition or a suboptimal solution by applying the Gaussian randomization method.
Finally, the $k$-th element of $\bm{p}^*$  is used to determine the power and phase allocation for node $k$, leading to improved accuracy in over-the-air computations. Moreover, we term the solution algorithm to Problem $\mathcal{P}_6$ as E-ChannelComp.

\begin{remark}
Due to the dependence of Problem $\mathcal{P}_6$ on the channel coefficients ($h_k$) in the case of a fast-fading channel, solving the optimization problem every time the channel changes can be computationally expensive. One alternative solution is separating the channel compensation and modulation adaptation steps. This means that optimization problem \eqref{eq:traceP-convex} only needs to be solved once for a given set of modulations and function $f$, similar to the approach in \eqref{eq:Penalty-convex}. However, this approach may be less power-efficient compared to the output of optimization problem \eqref{eq:traceP-convex}.
\end{remark}

In the next section, we describe the structure of the decoder of ChannelComp.  
\begin{figure*}
\centering
\subfigure[$f(\bm{x}) = \sum_{k=1}^Kx_k$]{
    \begin{tikzpicture} 
    \begin{axis}[
        xlabel={${\rm Real}(\bm{x})$},
        ylabel={${\rm Imag}(\bm{x})$},
        label style={font=\tiny},
        legend cell align={left},
        width=4.82cm,
        height=4.82cm,
        xmin=-7, xmax=7,
        ymin=-7, ymax=7,
        legend style={nodes={scale=0.65, transform shape}, at={(0.7,0.98)}}, 
        ymajorgrids=true,
        xmajorgrids=true,
        grid style=dashed,
        ticklabel style = {font=\tiny},
        grid=both,
        grid style={line width=.1pt, draw=gray!10},
        major grid style={line width=.2pt,draw=gray!30},
    ]
    \addplot[
        color=chestnut,
        mark=o,
        line width=1pt,
        mark size=2pt,
        ]
    table[x=X1,y=Y1]
    {Data/SumModulationK2Q3.dat};
    \addplot[
        color=airforceblue,
        mark=triangle,
        mark options = {rotate = 180},
        line width=0.5pt,
        mark size=2pt,
        ]
    table[x=X1,y=Y1]
    {Data/SumModulationK2Q3.dat};
    \legend{$\vec{x}_1$, $\vec{x}_2$};
    \end{axis}
\end{tikzpicture}
}
\subfigure[$f(\bm{x}) = \prod_{k=1}^Kx_k$]{
\label{fig:ConsQ4K2(b)}
   \begin{tikzpicture} 
    \begin{axis}[
        xlabel={${\rm Real}(\bm{x})$},
        ylabel={${\rm Imag}(\bm{x})$},
        label style={font=\tiny},
        width=4.82cm,
        height=4.82cm,
        xmin=-7, xmax=7,
        ymin=-7, ymax=7,
        ticklabel style = {font=\tiny},
        ymajorgrids=true,
        xmajorgrids=true,
        grid style=dashed,
        grid=both,
        grid style={line width=.1pt, draw=gray!10},
        major grid style={line width=.2pt,draw=gray!30},
    ]
    \addplot[
        color=chestnut,
        mark=o,
        line width=1pt,
        mark size=2pt,
        ]
    table[x=X1,y=Y1]
    {Data/ProdModK2Q3.dat};
    \addplot[
        color=airforceblue,
        mark=triangle,
        mark options = {rotate = 180},
        line width=0.5pt,
        mark size=2pt,
        ]
    table[x=X1,y=Y2]
    {Data/ProdModK2Q3.dat};
    \end{axis}
\end{tikzpicture}
}
\subfigure[$f(\bm{x}) = \max_kx_k$]{
    \begin{tikzpicture} 
    \begin{axis}[
        xlabel={${\rm Real}(\bm{x})$},
        ylabel={${\rm Imag}(\bm{x})$},
        label style={font=\tiny},
        width=4.82cm,
        height=4.82cm,
        xmin=-7, xmax=7,
        ymin=-7, ymax=7,
        ticklabel style = {font=\tiny},
        ymajorgrids=true,
        xmajorgrids=true,
        grid style=dashed,
        grid=both,
        grid style={line width=.1pt, draw=gray!10},
        major grid style={line width=.2pt,draw=gray!30},
    ]
    \addplot[
        color=chestnut,
        mark=o,
        line width=1pt,
        mark size=2pt,
        ]
    table[x=X1,y=Y1]
    {Data/MaxK2Q3.dat};
    \addplot[
        color=airforceblue,
        mark=triangle,
        mark options = {rotate = 180},
        line width=0.5pt,
        mark size=2pt,
        ]
    table[x=X2,y=Y2]
    {Data/MaxK2Q3.dat};
    \end{axis}
  
\end{tikzpicture}
}
\subfigure[$f(\bm{x}) = \frac{x_1+1}{x_2+1}$]{
   \begin{tikzpicture} 
    \begin{axis}[
        xlabel={${\rm Real}(\bm{x})$},
        ylabel={${\rm Imag}(\bm{x})$},
        label style={font=\tiny},
        width=4.82cm,
        height=4.82cm,
        xmin=-7, xmax=7,
        ymin=-7, ymax=7,
        ticklabel style = {font=\tiny},
        ymajorgrids=true,
        xmajorgrids=true,
        grid style=dashed,
        grid=both,
        grid style={line width=.1pt, draw=gray!10},
        major grid style={line width=.2pt,draw=gray!30},
    ]
    \addplot[
        color=chestnut,
        mark=o,
        line width=1pt,
        mark size=2pt,
        ]
    table[x=X1,y=Y1]
    {Data/FracModulationK2Q3.dat};
    \addplot[
        color=airforceblue,
        mark=triangle,
        mark options = {rotate = 180},
        line width=1pt,
        mark size=2pt,
        ]
    table[x=X2,y=Y2]
    {Data/FracModulationK2Q3.dat};
    \end{axis}
  20
\end{tikzpicture}
}
  \caption{ Constellation diagram of the modulation vector for $K=2$ nodes with $q=8$ ($3$ bits). The axes show the real and imaginary values corresponding to the baseband components.}
  \label{fig:ConsQ4K2}
   
\end{figure*}
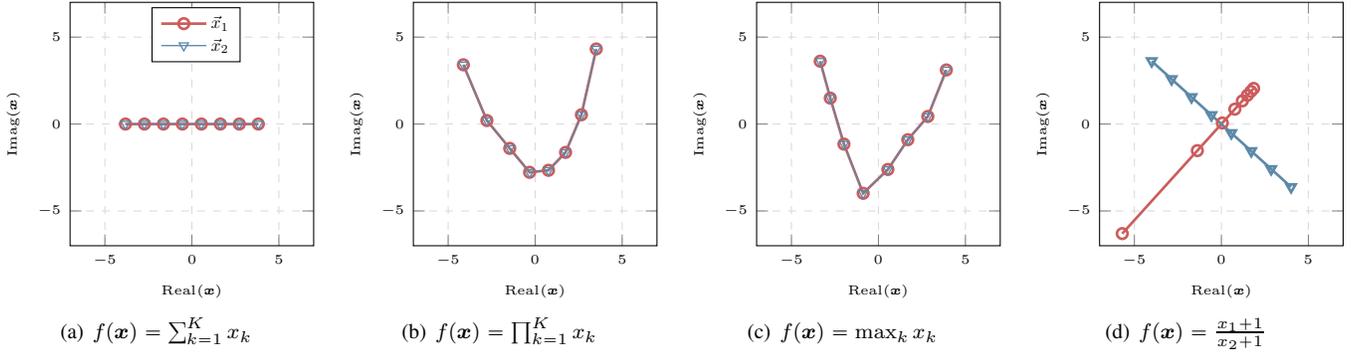

\subsection{Receiver Architecture in ChannelComp }\label{sec:Reciever}
After solving optimization problem $\mathcal{P}_3$ or $\mathcal{P}_6$ at the CP, the modulation vector $\bm{x}^*$ is determined. Then, the CP broadcasts the whole modulation vector $\bm{x}^*$'s to the nodes, and node $k$ uses $\bm{x}_k^*$ for the communication.  Afterward, the CP needs to determine the decoder $\mathscr{T}(\cdot)$ based on the obtained modulation vectors $\bm{x}_k^*$s. The decoder can be straightforwardly determined as long as the solution to either problems $\mathcal{P}_4$ or $\mathcal{P}_6$ is feasible. In particular, we can use the maximum likelihood estimator (MLE) for the decoder and design the decision boundaries based on the deformed constellation points at the CP server. Next, using the Tabular mapping $\mathscr{T}(\cdot)$ on the $\vec{y}(t)$ yields the desired value of the function $f$.  More precisely, let us define $\vec{g}_{i}:=\sum_{k=1}^K\vec{x}_k$, the constellation point corresponding to the function $f^{(i)}$. Then,  the problem is to find which $\vec{g}_{i}$'s values were transmitted while we have received $\vec{y}$. Hence, using the MLE estimator, we have 
\begin{align}
    \hat{f}^{(i)} = \underset{i}{\rm argmax}\Pr(\vec{y} | \vec{g}_{i}),    
\end{align}
where $\Pr(\vec{y} | \vec{g}_{i}) =  1/\sqrt{2\pi\sigma_z^2} \exp{\big[-\|\vec{y} -\vec{g}_i\|_2^2/{2\sigma_z^2}\big]}$ follows a Gaussian distribution. Next, taking logarithm results in the following expression 
\begin{align}
    \hat{f}^{(i)} = \underset{i}{\rm argmin} \| \vec{y} - \vec{g}_i\|_2^2.    
\end{align}
The last expression generates a Voronoi diagram of the set of all possible constellation points $\{\vec{g}_1,\ldots,\vec{g}_M\}$ with the corresponding Voronoi cells $\{\mathcal{V}_1,\ldots, \mathcal{V}_M\}$ \cite{SugiharaVoronoi1992}. Therefore, the desired value is given by 
$\hat{f} = \sum_{j=1}^{M}\mathscr{T}^{(j)}(\vec{y})$,
where 
\begin{align}
\label{eq:Tabularj}
      \mathscr{T}^{(j)}(\vec{g}):= \begin{cases}
            f^{(j)}, &    \vec{g}\in \mathcal{V}_j, \\
            0, & \text{otherwise}.
        \end{cases}
\end{align}
In the case where we cannot satisfy the computation condition in \eqref{eq:feasibility}, there may be points in which $f^{(i)}$ and $f^{(i+1)}$ have the same constellation point $\vec{g}_i$. In other words, these points have the same Voronoi cell, or $\mathcal{V}_i = \mathcal{V}_{i+1}$, and the decision rule can be replaced by their mean: 
\begin{align}
 \mathscr{T}^{(i)}(\vec{g})= \mathscr{T}^{(i+1)}(\vec{g}) = \begin{cases}
            \frac{f^{(i)}+f^{(i+1)}}{2},  &    \vec{g}\in \mathcal{V}_i, \mathcal{V}_{i+1}, \\
            0, & \text{otherwise}.
        \end{cases}
\end{align}
 
To illustrate,  consider one simple BPSK example in Figure~\ref{fig:BPSKExample} where there are $K=2$ nodes for computing the summation function, $f_1(x_1,x_2) = x_1 +x_2$ and product function, $f_2(x_1,x_2) = x_1 x_2$. Let $\mathscr{T}_{1}$ and $\mathscr{T}_{2}$ be the tabular maps  corresponding to functions $f_1$ and $f_2$, respectively. Then, for function $f_1$ the resultant constellation points are restricted to the set $\{-\Vec{2},\Vec{0},\Vec{2}\}$ , consequently, it gives the following  Voronoi cells, 
\begin{subequations}
\label{eq:Voronoi}
\begin{align}
    \mathcal{V}_1 & := \{\vec{g}~| {\rm Real}(\vec{g}) \leq  -1\},\\
    \mathcal{V}_2 & := \{\vec{g}~| -1 < {\rm Real}(\vec{g}) \leq 1\},\\
    \mathcal{V}_3 & := \{\vec{g}~|{\rm Real}(\vec{g})  > 1\}.
\end{align}
\end{subequations}
Hence, by substituting \eqref{eq:Voronoi} into \eqref{eq:Tabularj}, we can obtain $\mathscr{T}_1:= \sum_{i=1}^{3}\mathscr{T}^{(i)}_1$ for function $f_1$ as follows
\begin{align}
    \hat{f}_1 := \mathscr{T}_1(\vec{g}) = \begin{cases}
            0,  &    {\rm Real}(\vec{g}) \leq  -1, \\
            1,  &   -1 < {\rm Real}(\vec{g}) \leq 1 , \\
            2,  &    {\rm Real}(\vec{g})  > 1.
        \end{cases}
\end{align}
 Similarly for the $\mathscr{T}_{2}$, we have 
\begin{align}
    \mathscr{T}_2(\vec{g}) = \begin{cases}
            0,  &    {\rm Real}(\vec{g}) \leq  1 , \\
            1,  &    {\rm Real}(\vec{g})  > 1.
        \end{cases}
\end{align}

Note that the encoder and decoder in ChannelComp have a similar overhead compared to AirComp. This is because we only map the input and output using the modulation vectors obtained from Problem $\mathcal{P}_4$. The main complexity comes from solving optimization in \eqref{eq:traceX-convex}. This optimization must be done offline once before setting up the communication system. Note that thanks to the tailored modulation by ChannelComp, the latency of the communication does not depend on the number of nodes $K$; accordingly,  ChannelComp can handle a massive number of devices with a low latency communication system.

In the following section, we assess the performance of ChannelComp numerically.

\section{Numerical Experiments}\label{sec:Num}

In this section, we evaluate the performance of G-ChannelComp (solution to Problem $\mathcal{P}_3$), which designs the modulation among the nodes, and E-ChannelComp (solution to Problem $\mathcal{P}_6$), which adapts phase and power of modulations for the nodes, for several functions with different numbers of nodes. We compare the performance of G-ChannelComp and E-ChannelComp on the following two settings: $1.$  the standard digital transmission, with the computing method using the orthogonal frequency division multiple access (OFDMA), in which each node uses different frequency channels; and $2.$ to the AirComp method~\cite{goldenbaum2013harnessing}. In the last subsection, we repeat the comparison of ChannelComp with AirComp and OFDMA methods in the presence of fading.      
\begin{figure*}
\centering
\subfigure[$f(\bm{x}) = \sum_{k=1}^Kx_k$]{
    \begin{tikzpicture} 
    \begin{axis}[
        xlabel={${\rm Real}(\bm{x})$},
        ylabel={${\rm Imag}(\bm{x})$},
        label style={font=\tiny},
        legend cell align={left},
        width=4.82cm,
        height=4.82cm,
        xmin=-7, xmax=7,
        ymin=-7, ymax=7,
        legend style={nodes={scale=0.65, transform shape}, at={(0.45,0.98)}}, 
        ticklabel style = {font=\tiny},
        ymajorgrids=true,
        xmajorgrids=true,
        grid style=dashed,
        grid=both,
        grid style={line width=.1pt, draw=gray!10},
        major grid style={line width=.2pt,draw=gray!30},
    ]
    \addplot[
        color=chestnut,
        mark=o,
        line width=1pt,
        mark size=2pt,
        ]
    table[x=X1,y=Y1]
    {Data/Sim1.dat};
    \addplot[
        color=airforceblue,
        mark=triangle,
        mark options = {rotate = 180},
        line width=0.5pt,
        mark size=2pt,
        ]
    table[x=X2,y=Y2]
    {Data/Sim1.dat};
    \addplot[
        color=black,
        mark=triangle,
        line width=0.5pt,
        mark size=2pt,
        ]
    table[x=X4,y=Y4]
    {Data/Sim1.dat};
    \addplot[
        color=cadmiumorange,
        mark=star,
        line width=0.5pt,
        mark size=2pt,
        ]
    ttable[x=X3,y=Y2]
    {Data/Sim1.dat};
    \legend{$\vec{x}_1$, $\vec{x}_2$, $\vec{x}_3$, $\vec{x}_4$};
    \end{axis}
\end{tikzpicture}
}
\subfigure[$f(\bm{x}) = \prod_{k=1}^Kx_k$]{
\label{fig:ConsQ2K4(b)}
   \begin{tikzpicture} 
    \begin{axis}[
        xlabel={${\rm Real}(\bm{x})$},
        ylabel={${\rm Imag}(\bm{x})$},
        label style={font=\tiny},
        width=4.82cm,
        height=4.82cm,
        xmin=-7, xmax=7,
        ymin=-7, ymax=7,
        ticklabel style = {font=\tiny},
        ymajorgrids=true,
        xmajorgrids=true,
        grid style=dashed,
        grid=both,
        grid style={line width=.1pt, draw=gray!10},
        major grid style={line width=.2pt,draw=gray!30},
    ]
    \addplot[
        color=chestnut,
        mark=o,
        line width=1pt,
        mark size=2pt,
        ]
    table[x=X1,y=Y1]
    {Data/ProdModK4Q2.dat};
    \addplot[
        color=airforceblue,
        mark=triangle,
        mark options = {rotate = 180},
        line width=0.5pt,
        mark size=2pt,
        ]
    table[x=X2,y=Y2]
    {Data/ProdModK4Q2.dat};
    \addplot[
        color=black,
        mark=triangle,
        line width=0.5pt,
        mark size=2pt,
        ]
    table[x=X3,y=Y3]
    {Data/ProdModK4Q2.dat};
    \addplot[
        color=cadmiumorange,
        mark=star,
        line width=0.5pt,
        mark size=2pt,
        ]
    ttable[x=X4,y=Y4]
    {Data/ProdModK4Q2.dat};
    \end{axis}
  
\end{tikzpicture}
}
\subfigure[$f(\bm{x}) = \max_kx_k$]{
   \begin{tikzpicture} 
    \begin{axis}[
        xlabel={${\rm Real}(\bm{x})$},
        ylabel={${\rm Imag}(\bm{x})$},
        label style={font=\tiny},
        width=4.82cm,
        height=4.82cm,
        xmin=-7, xmax=7,
        ymin=-6, ymax=9,
        ticklabel style = {font=\tiny},
        ymajorgrids=true,
        xmajorgrids=true,
        grid style=dashed,
        grid=both,
        grid style={line width=.1pt, draw=gray!10},
        major grid style={line width=.2pt,draw=gray!30},
    ]
    \addplot[
        color=chestnut,
        mark=o,
        line width=1pt,
        mark size=2pt,
        ]
    table[x=X1,y=Y1]
    {Data/MaxModK4Q2.dat};
    \addplot[
        color=airforceblue,
        mark=triangle,
        mark options = {rotate = 180},
        line width=0.5pt,
        mark size=2pt,
        ]
    table[x=X2,y=Y2]
    {Data/MaxModK4Q2.dat};
    \addplot[
        color=black,
        mark=triangle,
        line width=0.5pt,
        mark size=2pt,
        ]
    table[x=X3,y=Y3]
    {Data/MaxModK4Q2.dat};
    \addplot[
        color=cadmiumorange,
        mark=star,
        line width=0.5pt,
        mark size=2pt,
        ]
    ttable[x=X4,y=Y4]
    {Data/MaxModK4Q2.dat};
    \end{axis}
  
\end{tikzpicture}
}
\subfigure[$f(\bm{x}) = \frac{x_1+1}{x_2+1} +\frac{x_4+3}{x_3+1} $]{
   \begin{tikzpicture} 
   \begin{scope}[spy using outlines={rectangle, magnification=2.5,
   width=1.2cm,height=1.2cm,connect spies}]
    \begin{axis}[
        xlabel={${\rm Real}(\bm{x})$},
        ylabel={${\rm Imag}(\bm{x})$},
        label style={font=\tiny},
        width=4.82cm,
        height=4.82cm,
        xmin=-4, xmax=7,
        ymin=-5, ymax=7,
        ymajorgrids=true,
        xmajorgrids=true,
        ticklabel style = {font=\tiny},
        grid style=dashed,
        grid=both,
        grid style={line width=.1pt, draw=gray!10},
        major grid style={line width=.2pt,draw=gray!30},
    ]
    \addplot[
        color=chestnut,
        mark=o,
        line width=0.5pt,
        mark size=1pt,
        ]
    table[x=X1,y=Y1]
    {Data/Sim2.dat};
    \addplot[
        color=airforceblue,
        mark=triangle,
        mark options = {rotate = 180},
        line width=0.5pt,
        mark size=1pt,
        ]
    table[x=X2,y=Y2]
    {Data/Sim2.dat};
    \addplot[
        color=black,
        mark=triangle,
        line width=0.5pt,
        mark size=1pt,
        ]
    table[x=X4,y=Y4]
    {Data/Sim2.dat};
    \addplot[
        color=cadmiumorange,
        mark=star,
        line width=0.5pt,
        mark size=1pt,
        ]
    ttable[x=X3,y=Y3]
    {Data/FracModK4Q2.dat};
     \path (0.3, 0.3) coordinate (X);
    \end{axis}
      \spy [black] on (X) in node (zoom) [left] at ([xshift=1.5cm,yshift=0.75cm]X);
    \end{scope}
\end{tikzpicture}
}
  \caption{ Constellation diagram of the modulation vector for $K=4$ nodes with $q=4$ ($2$ bits). The axes show the real and imaginary values corresponding to the baseband components.}
   \label{fig:ConsQ2K4}
\end{figure*}
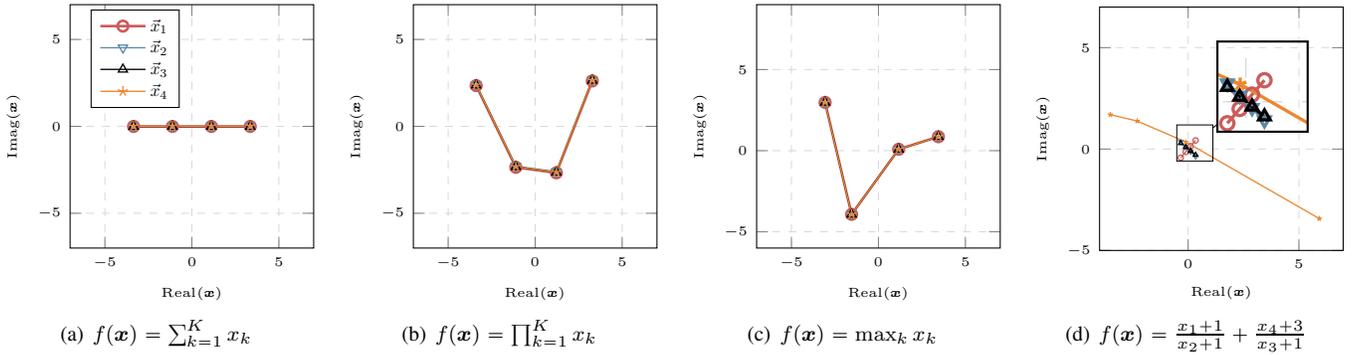

\subsection{Performance Evaluation of ChannelComp}

In the first numerical experiment, we solve optimization problem~\eqref{eq:traceX-convex}, i.e., (G-ChannelComp) Problem $\mathcal{P}_3$, for four different functions: summation, product, maximum, and fractional function. We consider different numbers of nodes and quantization values for each node, namely $K=2$ with $q=8$ and $K=4$ with $q=4$. For the case of $K=4$, we obtained an output $\bm{X}^*$, which is no longer a rank-one matrix. Hence, we solved optimization problem ~\eqref{eq:Penalty-convex}, i.e., Problem $\mathcal{P}_4$, with the penalty parameter to get a better approximation via Algorithm~\ref{alg:DC}. The  parameters of  Algorithm~\ref{alg:DC} are set as $\mu = 10^{-4}$ and $\delta = 10^{-3}$ in the simulations.  

The resultant modulation vectors are depicted in Figure~\ref{fig:ConsQ4K2}.  Figure~\ref{fig:ConsQ4K2} shows the resultant modulation vectors for the first three functions, i.e., summation, product, and maximum (symmetric functions), and the last fractional function results in different modulation vectors. In the case of the summation function, it turns out that the PAM modulation is the resultant modulation vector (see Figure~\ref{fig:ConsQ4K2}(a)). Specifically, $q=4$ is a misconception that QPSK  modulation is a more power-efficient alternative for computing the summation. However, this is not the case. It is shown that the PAM modulation is the optimal choice for computing the summation function, as it results in a lower computation error compared to other alternatives.

In Figure~\ref{fig:ConsQ2K4}, we still obtain similar modulations diagram results by increasing the number of nodes  to $K =4$ with $q = 4$. Furthermore, we observe that the modulation vector $\bm{x}$ of higher order, i.e., larger $q$, is an interpolated version of the lower order modulation. Similarly,  it shows that the lower order modulation can be obtained by sub-sampling from the higher order modulation instead of solving Problem $\mathcal{P}_3$. 

In the second experiment, we check the performance of E-ChannelComp for adapting the modulations as in problem~\eqref{eq:traceP-convex}, i.e., Problem $\mathcal{P}_6$, for the product function $f(\bm{x}) = \prod_{k=1}^Kx_k$ with $K=4$ using QPSK  modulation, i.e., $q=4$ and $\Vec{x} = [1,j,-1,j]^{\mathsf{T}}$. Moreover, to understand the effect of channel fading on the output of Problem $\mathcal{P}_6$, we repeat the experiment for fading channels whose coefficients are generated with normal Gaussian distribution $h_k\sim \mathcal{N}(0,1)$. 

\begin{figure}[!t]
\centering
\subfigure[Ideal channel]{\label{fig:QAM4(a)}
    \begin{tikzpicture} 
    \begin{axis}[
        xlabel={${\rm Real}(\bm{x})$},
        ylabel={${\rm Imag}(\bm{x})$},
        width=4.8cm,
        height=4.8cm,
        xmin=-1.2, xmax=1.2,
        ymin=-1.2, ymax=1.2,
        ticklabel style = {font=\footnotesize},
        label style={font=\tiny},
        ymajorgrids=true,
        xmajorgrids=true,
        grid style=dashed,
        grid=both,
        grid style={line width=.1pt, draw=gray!10},
        major grid style={line width=.2pt,draw=gray!30},
    ]
    \addplot[
        color=chestnut,
        mark=o,
        line width=1pt,
        mark size=3pt,
        ]
    table[x=X1,y=Y1]
    {Data/ModProd.dat};
    \addplot[
        color=airforceblue,
        mark=o,
        mark options = {rotate = 180},
        line width=1pt,
        mark size=3pt,
        ]
    table[x=X2,y=Y2]
    {Data/ModProd.dat};
    \addplot[
        color=black,
        mark=triangle,
        line width=1pt,
        mark size=3pt,
        ]
    table[x=X3,y=Y3]
    {Data/ModProd.dat};
    \addplot[
        color=cadmiumorange,
        mark=star,
        line width=1pt,
        dashed,
        mark size=3pt,
        ]
    ttable[x=X4,y=Y4]
    {Data/ModProd.dat};
    \end{axis}
\end{tikzpicture}
}\subfigure[Fading channel]{\label{fig:QAM4(b)}
    \begin{tikzpicture} 
    \begin{axis}[
        xlabel={${\rm Real}(\bm{x})$},
        ylabel={${\rm Imag}(\bm{x})$},
        width=4.8cm,
        height=4.8cm,
        legend style={nodes={scale=0.55, transform shape}, at={(0.38,0.98)}}, 
        ticklabel style = {font=\footnotesize},
        label style={font=\tiny},
        ymajorgrids=true,
        xmajorgrids=true,
        grid style=dashed,
        grid=both,
        grid style={line width=.1pt, draw=gray!10},
        major grid style={line width=.2pt,draw=gray!30},
    ]
    \addplot[
        color=chestnut,
        mark=o,
        line width=1pt,
        mark size=3pt,
        ]
    table[x=X1,y=Y1]
    {Data/ModRandomProd.dat};
    \addplot[
        color=airforceblue,
        mark=o,
        mark options = {rotate = 180},
        line width=1pt,
        mark size=3pt,
        ]
    table[x=X2,y=Y2]
    {Data/ModRandomProd.dat};
    \addplot[
        color=black,
        mark=triangle,
        line width=1pt,
        mark size=3pt,
        ]
    table[x=X3,y=Y3]
    {Data/ModRandomProd.dat};
    \addplot[
        color=cadmiumorange,
        mark=star,
        line width=1pt,
        dashed,
        mark size=3pt,
        ]
    ttable[x=X4,y=Y4]
    {Data/ModRandomProd.dat};
    \legend{$\vec{x}_1$, $\vec{x}_2$, $\vec{x}_3$, $\vec{x}_4$};
    \end{axis}
\end{tikzpicture}
}
  \caption{ The resultant modulation vector of \eqref{eq:traceP-convex} for computing the product function $f(\bm{x}) =\prod_{k}x_k$ with $K=4$ nodes and quantization $q=4$, when the nodes use QPSK modulation. Figures \ref{fig:QAM4(a)} and \ref{fig:QAM4(b)} show the modulation each node should use, in the case of ideal and fading channels, respectively. The x- and y-axis of the plot illustrate the real and imaginary parts of the baseband components, respectively.}
   \label{fig:QAM4}
\end{figure}
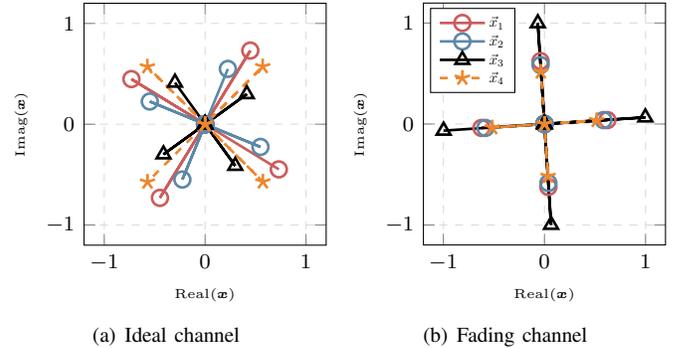

Figure~\ref{fig:QAM4} shows the resultant modulations whose power and phases are given by the solution to problem \eqref{eq:traceP-convex} for the ideal and fading channels in  Figures~\ref{fig:QAM4(a)} and \ref{fig:QAM4(b)}, respectively. We observe that if the fading channel is not considered, E-ChannelComp needs to adapt the power and phase of the modulations to avoid that destructive overlapping occurring over the MAC. Conversely, in the presence of fading, the channel coefficients can alter the powers and phase of the modulations obtained at the receiver, resulting in a more power-efficient manner than standard OFDMA over the MAC. Specifically, in Figure~\ref{fig:QAM4(a)}, node $2$ needs to reduce its power to satisfy the computation constraints in  \eqref{eq:traceP-convex}. At the same time, a low value of $h_2$ suggests not compensating the channel effect and allocating more power to this node.

\begin{figure*}
\centering
\subfigure[]{
 \label{fig:QunNMSE(a)}
    \begin{tikzpicture} 
    \begin{axis}[
        xlabel={SNR(dB)},
        ylabel={NMSE},
        label style={font=\scriptsize},
        legend cell align={left},
        tick label style={font=\scriptsize} , 
        width=8cm,
        height=7.5cm,
        xmin=-5, xmax=27,
        ymin=1e-3, ymax=1e2,
        ymode = log,
       legend style={nodes={scale=0.65, transform shape}, at={(0.98,0.98)}}, 
        ymajorgrids=true,
        xmajorgrids=true,
        grid style=dashed,
        grid=both,
        grid style={line width=.1pt, draw=gray!15},
        major grid style={line width=.2pt,draw=gray!40},
    ]
    \addplot[
        color=bleudefrance,
        mark=square,
        line width=1pt,
        mark size=2pt,
        ]
    table[x=SNR,y=ChannelComp]
    {Data/MSEQError.dat};
     \addplot[
        color=cadmiumorange,
        mark=square,
        line width=1pt,
        mark size=2pt,
        ]
    table[x=SNR,y=OAT]
    {Data/MSEQError.dat};
    \addplot[
        color=cadmiumgreen,
        mark=square,
        line width=1pt,
        mark size=2pt,
        ]
    table[x=SNR,y=OFDM]
    {Data/MSEQError.dat};
    \addplot[
        color=bleudefrance,
        mark=o,
        mark options = {rotate = 180},
        line width=1pt,
        mark size=2pt,
        ]
    table[x=SNR,y=ChannelComp]
    {Data/MSEProdcutError.dat};
     \addplot[
        color=cadmiumorange,
         mark=o,
        mark options = {rotate = 180},
        line width=1pt,
        mark size=2pt,
        ]
    table[x=SNR,y=OAT]
    {Data/MSEProdcutError.dat};
    \addplot[
        color=cadmiumgreen,
        mark=o,
         mark options = {rotate = 180},
        line width=1pt,
        mark size=2pt,
        ]
    table[x=SNR,y=OFDM]
    {Data/MSEProdcutError.dat};
    \legend{G-ChannelComp $\sum $,AirComp$\sum $, OFDMA$\sum $,G-ChannelComp $\prod $,AirComp$\prod $,OFDMA$\prod $};
    \end{axis}
\end{tikzpicture}
}   \subfigure[]{
 \label{fig:QunNMSE(b)}
  \begin{tikzpicture} 
    \begin{axis}[
        xlabel={SNR(dB)},
        ylabel={NMSE},
        label style={font=\scriptsize},
        legend cell align={left},
        tick label style={font=\scriptsize} , 
        width=8cm,
        height=7.5cm,
        xmin=-5, xmax=27,
        ymin=1e-3, ymax=1e2,
        ymode = log,
       legend style={nodes={scale=0.55, transform shape}, at={(0.98,0.98)}}, 
        ymajorgrids=true,
        xmajorgrids=true,
        grid style=dashed,
        grid=both,
        grid style={line width=.1pt, draw=gray!15},
        major grid style={line width=.2pt,draw=gray!40},
    ]
    \addplot[
        color=bleudefrance,
        mark=square,
        line width=1pt,
        mark size=2pt,
        ]
    table[x=SNR,y=ChannelComp]
    {Data/MSEQunatizedErrorMax.dat};
     \addplot[
        color=cadmiumorange,
         mark=square,
        line width=1pt,
        mark size=2pt,
        ]
    table[x=SNR,y=OAT]
    {Data/MSEQunatizedErrorMax.dat};
   \addplot[
        color=cadmiumgreen,
        mark=square,
        line width=1pt,
        mark size=2pt,
        ]
    table[x=SNR,y=OFDM]
    {Data/MSEQunatizedErrorMax.dat};
    \addplot[
        color=bleudefrance,
        mark=o,
        line width=1pt,
        mark size=2pt,
        ]
    table[x=SNR,y=ChannelComp]
    {Data/MSEQuantizedQUAD.dat};
     \addplot[
        color=cadmiumorange,
         mark=o,
        line width=1pt,
        mark size=2pt,
        ]
    table[x=SNR,y=OAT]
    {Data/MSEQuantizedQUAD.dat};
    \addplot[
        color=cadmiumgreen,
        mark=o,
        line width=1pt,
        mark size=2pt,
        ]
    table[x=SNR,y=OFDM]
    {Data/MSEQuantizedQUAD.dat};
    \legend{G-ChannelComp  $\max $,AirComp $\max $,OFDMA $\max $, G-ChannelComp  $\bm{x}^2$,AirComp $\bm{x}^2$,OFDMA $\bm{x}^2 $};
    \end{axis}
\end{tikzpicture}

}
  \caption{Performance comparison between G-ChannelComp, AirComp, and OFDMA in terms of NMSE error averaged over $N_s =100$, when values of the function to be computed are originally quantized. Such input values are the input value $x_k=\{0,1,\ldots,7\}$  and the desired functions are $f_1 = \sum_{k=1}^4x_k$,$f_2 = \prod_{k=1}^4x_k$, $f_3 = \max_{k}x_k$, and $f_4 = \sum_{k=1}^4x_k^2$.} 
  \label{fig:QunNMSE}
   
\end{figure*}
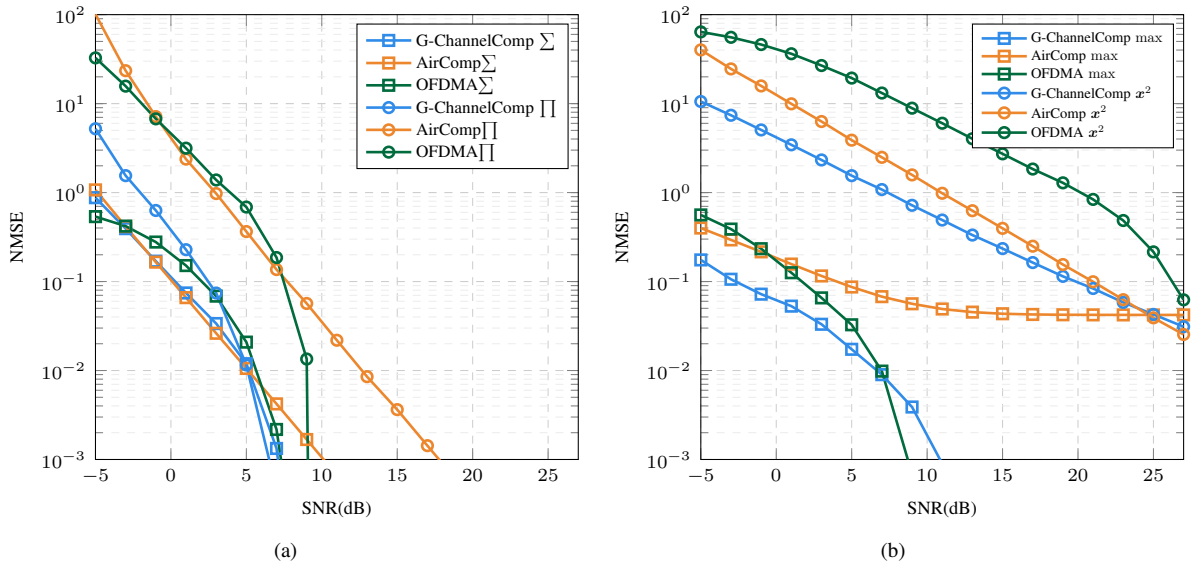

\subsection{Comparison to AirComp}

We compare G-ChannelComp to OFDMA and AirComp, which is analog and mostly intended for summation and product functions. Hence, we compare G-ChannelComp, AirComp, and OFDMA using these functions in different scenarios where the input data has either continuous or discrete values. Specifically, we consider the function's input values originally quantized in the first scenario.  Thus, the input values are naturally associated with digital values modulated over the AWGN channel. In other words, we consider the summation function $f = \sum_{k=1}^{4}x_k$, where $x_k\in \{0,1,2, \ldots, 8\}$ over a network with $K=4$ nodes.   For a fair comparison, the peak of modulated signals' amplitude of G-ChannelComp is set to be the same as the maximum power for AirComp, which is the maximum value of input value or $8$.  Furthermore, to characterize the computation error in G-ChannelComp, AirComp, and OFDMA, we use the normalized mean square error (NMSE) metric, which is defined as 
${\rm NMSE}:= \sum_{j=1}^{N_s}|f^{(i)} - \hat{f}_j^{(i)}|^2/{N_s|f^{(i)}|},$
where $N_s$ denotes the number of Monte Carlo trials, $f^{(i)}$ denotes the value of the desired function we wish to compute, and $\hat{f}_i^{(j)}$ is the $j$-th estimated value of $f^{(i)}$ for $j\in [N_s]$.

Figure~\ref{fig:QunNMSE(a)} shows the NMSE  for different signal-to-noise ratios (SNRs), which is defined as ${\rm SNR}:=20\log(\|\bm{x}\|_2/\sigma_z)$.  We note that G-ChannelComp has a similar performance to AirComp for computing the summation function at the low SNR (less than $5$ dB), but as the SNR increases, G-ChannelComp outperforms AirComp. For the product function,  G-ChannelComp shows approximately a $10$ dB improvement compared to the other methods, even in low SNR scenarios. We repeat this experiment for the maximum  $f = \max_{k}x_K$ and quadratic $f = \sum_{k=1}^Kx_k^2$ functions in Figure~\ref{fig:QunNMSE(b)}.

We observe that G-ChannelComp exhibits superior performance compared to the other methods for both functions under examination, thanks to the constructive overlapping of constellation points of all the nodes. Furthermore, it is notable that AirComp cannot accurately compute the maximum function even in low SNR scenarios, likely due to its approximation techniques applied to the maximum function \footnote{ AirComp approximates the maximum using the log sum function. Indeed, $f(x_1,\ldots,x_K) = \log{(\sum_{k=1}^K\exp{(x_k)})} \approx \max_kx_{k}$ \cite{sahin2022survey} where we need to set $\mathscr{E}(x_k) :=\exp{(x_k)}$ and $\mathscr{T}(y) := \log(y)$ as encoders and Tabular mapping, respectively. 
However, ChannelComp computes the exact value for such a function in a noise-free communication by solving optimization in Problem~$\mathcal{P}_3$.}. For the quadratic function, the results suggest that G-ChannelComp outperforms both other methods across a wide range of SNR, with a particularly noticeable advantage in low SNR scenarios.

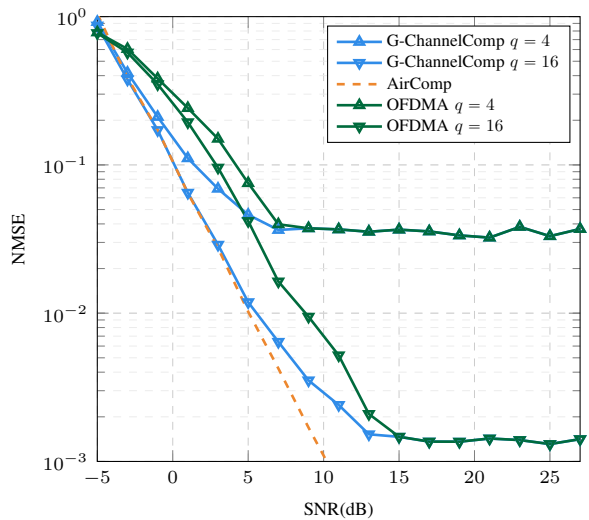
\begin{figure}
\centering
    \begin{tikzpicture} 
    \begin{axis}[
        xlabel={SNR(dB)},
        ylabel={NMSE},
        label style={font=\scriptsize},
        legend cell align={left},
        tick label style={font=\scriptsize} , 
        width=8cm,
        height=7.5cm,
        xmin=-5, xmax=27,
        ymin=1e-3, ymax=1,
        ymode = log,
       legend style={nodes={scale=0.65, transform shape}, at={(0.98,0.98)}}, 
        ymajorgrids=true,
        xmajorgrids=true,
        grid style=dashed,
        grid=both,
        grid style={line width=.1pt, draw=gray!15},
        major grid style={line width=.2pt,draw=gray!40},
    ]
    \addplot[
        color=bleudefrance,
        mark=triangle,
        line width=1pt,
        mark size=2pt,
        ]
    table[x=SNR,y=ChannelCompQ1]
    {Data/MSEQWError.dat};
    \addplot[
        color=bleudefrance,
        mark=triangle,
        mark options = {rotate = 180},
        line width=1pt,
        mark size=2pt,
        ]
    table[x=SNR,y=ChannelCompQ2]
    {Data/MSEQWError.dat};
  \addplot[
        color=cadmiumorange,
        dashed,
        line width=1pt,
        ]
    table[x=SNR,y=OAT]
    {Data/MSEQWError.dat};
     \addplot[
        color=cadmiumgreen,
        mark=triangle,
        line width=1pt,
        mark size=2pt,
        ]
    table[x=SNR,y=OFDMQ1]
    {Data/MSEQWError.dat};
     \addplot[
    color=cadmiumgreen,
    mark=triangle,
    mark options = {rotate = 180},
    line width=1pt,
    mark size=2pt,
    ]
table[x=SNR,y=OFDMQ2]
{Data/MSEQWError.dat};
    \legend{G-ChannelComp $q =4$,G-ChannelComp $q = 16$, AirComp, OFDMA $q=4$,OFDMA $q=16$};
    \end{axis}
\end{tikzpicture}

  \caption{Performance comparison between G-ChannelComp, AirComp, and OFDMA in terms of NMSE error for computing the summation function  $f = \sum_{k=1}^4x_k$ with the continuous input values $x_k$ in the interval $[0,7]$, averaged over $N_s =100$. }
  \label{fig:QunWithNMSE}
   
\end{figure}

In the next experiment reported in Figure~\ref{fig:QunWithNMSE}, we generate uniform random values as continuous numbers, $x_k \sim [0,7]$. Afterward, for G-ChannelComp, these values are quantized with $q=4$ and $q=16$ levels (or equivalently $2$ and $4$ bits) and transmitted over the MAC. Figure~\ref{fig:QunWithNMSE} shows that the performance of G-ChannelComp is saturated by quantization noise level for low noise cases. However, increasing the number of bits can mitigate this issue. The improved performance observed for AirComp can be attributed to the fact that it does not employ quantization, thus resulting in a transmitted signal with a lower noise level and preserving the signal's quality. Moreover, when the noise variance is very high (SNR less than $-3$ dB), OFDMA performs better than G-ChannelComp in the summation function, which comes from the input values having limited domain \footnote{When the variance of the noise is very high, we only observe the value of the boundary of the input domain with high probability. In fact, for OFDMA, with high probability, the estimation is either $0$ or $7$. As a result, the estimation of the summation is more likely around $14$ for $K = 4$ nodes. However, for G-ChannelComp, we directly compute the summation. Accordingly, the estimation of the summation would be either $0$ or $28$ with a high probability, which leads to more error. } ($x_k$s are between $0$ and $7$).  

\begin{figure}
\subfigure[]{
\label{fig:Modulation-QAM(a)}
    \begin{tikzpicture} 
    \begin{axis}[
        xlabel={SNR(dB)},
        ylabel={NMSE},
        label style={font=\scriptsize},
        legend cell align={left},
        tick label style={font=\scriptsize} , 
        width=8cm,
        height=7.5cm,
        xmin=-5, xmax=27,
        ymin=1e-2, ymax=10,
        ymode = log,
       legend style={nodes={scale=0.45, transform shape}, at={(0.98,0.98)}}, 
        ymajorgrids=true,
        xmajorgrids=true,
        grid style=dashed,
        grid=both,
        grid style={line width=.1pt, draw=gray!15},
        major grid style={line width=.2pt,draw=gray!40},
    ]
    \addplot[
        color=bleudefrance,
        mark=o,
        mark options = {rotate = 180},
        line width=1pt,
        mark size=2pt,
        ]
    table[x=SNR,y=ChannelComp]
    {Data_Standard/QPSKSUM.dat};
        \addplot[
        color=cadmiumgreen,
        mark=o,
        mark options = {rotate = 180},
        line width=1pt,
        mark size=2pt,
        ]
    table[x=SNR,y=OFDM]
    {Data_Standard/QPSKSUM.dat};
    \addplot[
        color=bleudefrance,
        mark=star,
        line width=1pt,
        mark size=2pt,
        ]
    table[x=SNR,y=ChannelComp]
    {Data_Standard/MSEQAM16SUM.dat};
        \addplot[
        color=cadmiumgreen,
        mark=star,
        line width=1pt,
        mark size=2pt,
        ]
    table[x=SNR,y=OFDM]
    {Data_Standard/MSEQAM16SUM.dat};
    \addplot[
        color=bleudefrance,
        mark=triangle,
        line width=1pt,
        mark size=2pt,
        ]
    table[x=SNR,y=ChannelComp]
    {Data_Standard/MSEQAM64SUM.dat};
        \addplot[
        color=cadmiumgreen,
        mark=triangle,
        line width=1pt,
        mark size=2pt,
        ]
    table[x=SNR,y=OFDM]
    {Data_Standard/MSEQAM64SUM.dat};
        \addplot[
        color=bleudefrance,
        mark=square,
        line width=1pt,
        mark size=2pt,
        ]
    table[x=SNR,y=ChannelComp]
    {Data_Standard/MSEQAM256SUM.dat};
    \addplot[
        color=cadmiumgreen,
        mark=square,
        line width=1pt,
        mark size=2pt,
        ]
    table[x=SNR,y=OFDM]
    {Data_Standard/MSEQAM256SUM.dat};
    \legend{E-ChannelComp QAM 4, OFDMA QAM 4,E-ChannelComp QAM 16,OFDMA QAM 16,E-ChannelComp QAM 64,OFDMA QAM 64, E-ChannelComp QAM 256, OFDMA QAM 256 };
    \end{axis}
\end{tikzpicture}
}
\subfigure[]{
\label{fig:Modulation-QAM(b)}
 \begin{tikzpicture} 
    \begin{axis}[
        xlabel={Number of nodes $[K]$},
        ylabel={NMSE},
        label style={font=\scriptsize},
        legend cell align={left},
        tick label style={font=\scriptsize} , 
        width=8cm,
        height=7.5cm,
        xmin=2, xmax=64,
        ymin=1e-2, ymax=1e2,
        ymode = log,
       legend style={nodes={scale=0.6, transform shape}, at={(0.98,0.98)}}, 
        ymajorgrids=true,
        xmajorgrids=true,
        grid style=dashed,
        grid=both,
        grid style={line width=.1pt, draw=gray!15},
        major grid style={line width=.2pt,draw=gray!40},
    ]
   \addplot[
        color=amethyst,
        mark=star,
        line width=0.75pt,
        mark size=1.75pt,
        ]
    table[x=nodes,y=QAM64]
    {Data/MSEQAMnodes.dat};
    \addplot[
        color=antiquebrass,
        mark=star,
        line width=0.75pt,
        mark size=1.75pt,
        ]
    table[x=nodes,y=QAM256]
    {Data/MSEQAMnodes.dat};
    \addplot[
        color=amaranth,
        mark=star,
        mark options = {rotate = 180},
        line width=0.75pt,
        mark size=1.75pt,
        ]
    table[x=nodes,y=QAM1024]
    {Data/MSEQAMnodes.dat};
    \legend{QAM $64$, QAM $256$, QAM $1024$};
    \end{axis}
\end{tikzpicture}

}
  \caption{Performance of adapting modulation using E-ChannelComp over the MAC for the QAM modulation.  The function is  summation $f = \sum_{k=1}^Kx_k$ over $N_s = 1000$  samples randomly selected from $\mathcal{R}_f$. Figure \ref{fig:Modulation-QAM(a)} shows the performance of the E-ChannelComp compared to OFDMA for $k=10$ nodes. Figure \ref{fig:Modulation-QAM(b)} shows the performance of the E-ChannelComp for different numbers of nodes $K$ when the ${\rm SNR} = 10$ dB. Increasing the number of nodes decreases the error thanks to the constructive interference.  }
  \label{fig:Modulation-QAM}
   
\end{figure}
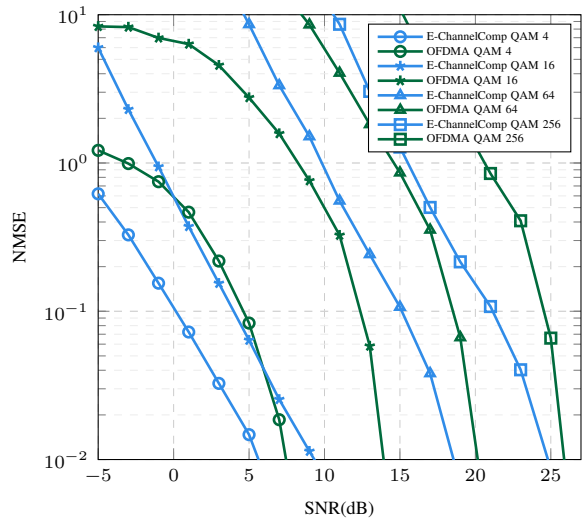
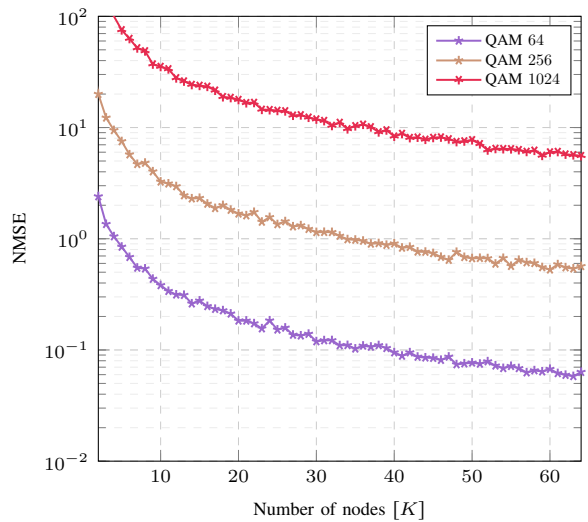

The results of this subsection show that the designed modulation by G-ChannelComp outperforms the other approaches, specifically for non-summation functions. G-ChannelComp and AirComp obtain similar performance for the summation function because AirComp can be considered a special case of G-ChannelComp (Remark \ref{rem:P2}).  

\subsection{E-ChannelComp with QAM Modulations}

In this subsection,  we analyze the performance of E-ChannelComp for QAM modulation with multiple orders and compare it to OFDMA. We consider a network of $K=10$ nodes for computing the summation function $f = \sum_kx_k$.  
Figure~\ref{fig:Modulation-QAM(a)} shows the NMSE performance of \Comp and OFDMA using QAM $4$, QAM $16$, QAM $64$, QAM $256$ modulations' depicted by $N_s =1000$ samples randomly selected from the range of the function $f(x_1,\ldots,x_K)$, i.e., $\mathcal{R}_f$. The results show that E-ChannelComp performs better than OFDMA, thanks to constructive overlaps of E-ChannelComp, where multiple signals are transmitted simultaneously and overlap in the frequency domain, resulting in a stronger signal than the individual signals in OFDMA. In addition, increasing the order of modulation results in lower performance for both E-ChannelComp and OFDMA due to reaching the channel's capacity.  

\begin{figure}
\subfigure[$f = \prod x_k$]{
\label{fig:Fading(a)}
    \begin{tikzpicture} 
    \begin{axis}[
        xlabel={SNR(dB)},
        ylabel={NMSE},
        label style={font=\scriptsize},
        legend cell align={left},
        tick label style={font=\scriptsize} , 
        width=8cm,
        height=7.5cm,
        xmin=-5, xmax=25,
        ymin=1e-2, ymax=1e3,
        ymode = log,
       legend style={nodes={scale=0.6, transform shape}, at={(0.98,0.98)}}, 
        ymajorgrids=true,
        xmajorgrids=true,
        grid style=dashed,
        grid=both,
        grid style={line width=.1pt, draw=gray!15},
        major grid style={line width=.2pt,draw=gray!40},
    ]
   \addplot[
        color=bleudefrance,
        mark=square,
        line width=1pt,
        mark size=2pt,
        ]
    table[x=SNR,y=ChannelComp]
    {Data/MSEFading.dat};
    \addplot[
        color=cadmiumorange,
        mark=square,
        mark options = {rotate = 180},
        line width=1pt,
        mark size=2pt,
        ]
    table[x=SNR,y=OAT]
    {Data/MSEFading.dat};
    \addplot[
        color=cadmiumgreen,
        mark=square,
        line width=1pt,
        mark size=2pt,
        ]
    table[x=SNR,y=OFDM]
    {Data/MSEFading.dat};
   \addplot[
        color=bleudefrance,
        mark=o,
        line width=1pt,
        mark size=2pt,
        ]
    table[x=SNR,y=ChannelComp]
    {Data/MSEFading2.dat};
    \addplot[
        color=cadmiumorange,
        mark=o,
        mark options = {rotate = 180},
        line width=1pt,
        mark size=2pt,
        ]
    table[x=SNR,y=OAT]
    {Data/MSEFading2.dat};
    \addplot[
        color=cadmiumgreen,
        mark=o,
        line width=1pt,
        mark size=2pt,
        ]
    table[x=SNR,y=OFDM]
    {Data/MSEFading2.dat};
    \legend{E-ChannelComp $K=4$,AirComp $K=4$,OFDMA $K=4$, E-ChannelComp $K=2$,AirComp $K=2$,OFDMA $K=2$};
    \end{axis}
\end{tikzpicture}}
\subfigure[$f = \max x_k$]{
\label{fig:Fading(b)}
    \begin{tikzpicture} 
    \begin{axis}[
        xlabel={SNR(dB)},
        ylabel={NMSE},
        label style={font=\scriptsize},
        legend cell align={left},
        tick label style={font=\scriptsize} , 
        width=8cm,
        height=7.5cm,
        xmin=-5, xmax=25,
        ymin=1e-3, ymax=1e1,
        ymode = log,
       legend style={nodes={scale=0.6, transform shape}, at={(0.98,0.98)}}, 
        ymajorgrids=true,
        xmajorgrids=true,
        grid style=dashed,
        grid=both,
        grid style={line width=.1pt, draw=gray!15},
        major grid style={line width=.2pt,draw=gray!40},
    ]
   \addplot[
        color=bleudefrance,
        mark=square,
        line width=1pt,
        mark size=2pt,
        ]
    table[x=SNR,y=ChannelComp]
    {Data/MSEFadingMAX.dat};
    \addplot[
        color=cadmiumorange,
        mark=square,
        mark options = {rotate = 180},
        line width=1pt,
        mark size=2pt,
        ]
    table[x=SNR,y=OAT]
    {Data/MSEFadingMAX.dat};
    \addplot[
        color=cadmiumgreen,
        mark=square,
        line width=1pt,
        mark size=2pt,
        ]
    table[x=SNR,y=OFDM]
    {Data/MSEFadingMAX.dat};
   \addplot[
        color=bleudefrance,
        mark=o,
        line width=1pt,
        mark size=2pt,
        ]
    table[x=SNR,y=ChannelComp]
    {Data/MSEFadingMAXK2.dat};
    \addplot[
        color=cadmiumorange,
        mark=o,
        mark options = {rotate = 180},
        line width=1pt,
        mark size=2pt,
        ]
    table[x=SNR,y=OAT]
    {Data/MSEFadingMAXK2.dat};
    \addplot[
        color=cadmiumgreen,
        mark=o,
        line width=1pt,
        mark size=2pt,
        ]
    table[x=SNR,y=OFDM]
    {Data/MSEFadingMAXK2.dat};
    \legend{E-ChannelComp $K=4$,AirComp $K=4$,OFDMA $K=4$, E-ChannelComp $K=2$,AirComp $K=2$,OFDMA $K=2$};
    \end{axis}
\end{tikzpicture}
}
  \caption{Performance comparison between E-ChannelComp, AirComp, and OFDMA in terms of NMSE versus the SNR values in the presence of fading channels. We considered $K=2$, and $4$ nodes, and computing the product and the max function using QPSK in \ref{fig:Fading(a)} and \ref{fig:Fading(b)}, respectively. The channel coefficients are generated by the Gaussian distribution, i.e., $h_k\sim \mathcal{N}(0,1)$.}
  \label{fig:Fading}
   
\end{figure}
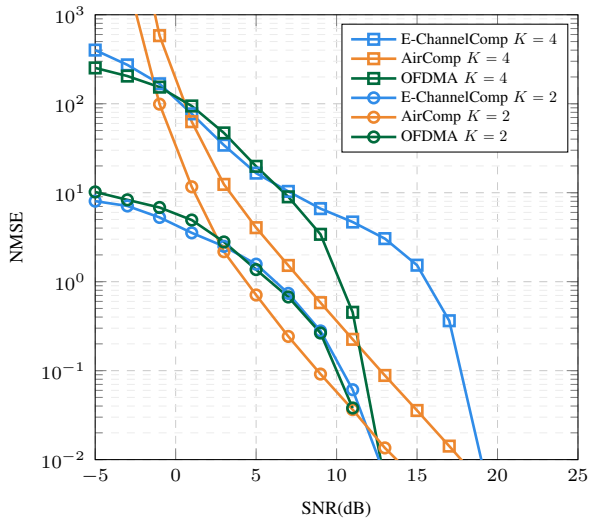
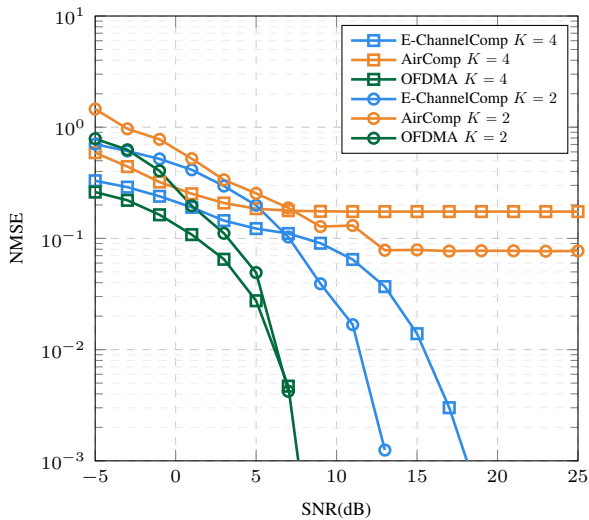

In Figure \ref{fig:Modulation-QAM(b)}, we analyze the effect of the number of nodes $K$ on the performance of \Comp using QAM $64,256$ and $1024$ modulation in an AWGN channel with ${\rm SNR} = 10$ dB for computing the summation function. We note  that increasing the number of nodes decreases the NMSE error, which comes from constructive interference of node signals.

For the last experiment, we analyze the effect of fading channels on the performance using QPSK modulation for computing the product and max functions of $K=4$ nodes and $K=2$, i.e.,  $f = \prod_{k=1}^{4}x_k$ and $f = \max_{k}x_k$. Figure \ref{fig:Fading(a)}  shows that \Comp  can achieve better NMSE performance than AirComp in high SNR scenarios, while it has a similar performance to OFDMA in  low SNR scenarios.  We note that the deterioration of E-ChannelCompp performance in this experiment comes from the fact that  QPSK modulation is not the optimal modulation for computing the product function (see Figure \ref{fig:ConsQ2K4(b)}). Similarly, for the max function in Figure \ref{fig:Fading(b)}, E-ChannelComp performs better than AirComp for $K=2$ and $K=4$ using the same communication resources. Also, OFDMA shows the best performance here, using more bandwidths than the other methods. 

We observe that E-ChannelComp enables existing modulation to compute functions over the MAC; however, the performance drops compared to the designed modulation by G-ChannelComp.

\section{Conclusion}\label{sec:conclusion}
In this study, we presented a novel over-the-air computation principle and method called ChannelComp, which utilizes digital or analog modulation to compute functions over multiple access channels. Our proposed method can compute a much broader class of functions compared to the AirComp method, which is restricted to analog modulations.  ChannelComp can handle massive devices simultaneously, ensuring strict computation time constraints.

We also proposed an offline modulation selection method based on optimizing a feasibility function and a tabular mapping capable of computing functions using digital modulation. We showed how to adapt existing digital modulation schemes to compute the desired function over the MAC. The Simulation results showed that ChannelComp outperforms analog AirComp and OFDMA methods in terms of normalized mean squared error with a notable improvement of around $10$ dB in computation error.

There are numerous potential avenues for further investigation of ChannelComp, including, but not limited to, the following: 
\begin{itemize}
    \item \textbf{Robust solution for stochastic fading channel:}
    we plan to extend ChannelComp to accommodate general functions through different (digital and analog) modulations for each node and evaluate the impact of stochastic fading.
    \item \textbf{Other optimization perspectives:} we solved a feasibility problem to find modulation vectors for a valid computation over-the-air. One can use the proposed computation to optimize the other objectives, e.g., minimizing the maximum or average computation error.  
    \item \textbf{Time variant systems:} Both G-ChannelComp and E-ChannelComp can be extended for designing or adapting the modulation vector for a general time-varying system model. As a result, it enables either computing a sequence of functions over time or stabilizing the computation against the channel changes. 
    \item \textbf{MIMO extension for matrix computation:} we intend to expand the current single narrowband antenna system at the transmitters and receiver to broadband multiple-input and multiple-output systems, enabling vector-based computations for several applications, such as matrix computation or federated learning.
    \item \textbf{Sequential transmission:} we plan to introduce repetitions in the transmissions and look at the sequence of received symbols across a trellis diagram to tradeoff modulation complexity with computation time and computation precision. 
    \item \textbf{Machine learning applications:} we aim to demonstrate that ChannelComp has the potential to enhance applications such as federated edge learning significantly.
\end{itemize}


 \appendix
 
 \subsection{Proof of Proposition \ref{pro:Rang} }\label{proof:PropRange}
 
 In the case where all nodes use identical modulation $\mathscr{E}_k=\mathscr{E}$ for $k\in [K]$, the summation over-the-air can be considered as a symmetric function $\mathscr{G}(\cdot)= \sum_{k=1}^K\mathscr{E}(\cdot)$.  Consequently, to compute the upper bound, the outputs of the function need to be as much as possible distinguishable, which means that all the outputs of $\mathscr{E}(\cdot)$ are distinct. Therefore, the range of $\mathscr{E}(x)$ would be $q$  for $x\in \{0,1,\ldots,q-1\}$. Next, we need to count the number of possible distinct outputs for the function $\mathscr{G}(x_1,\ldots,x_K)=\sum_{k=1}^K\mathscr{E}(x_k)$. Each $x_k$ has $q$ distinct values, and we need to count the number of ways to select one level of $x_k$ such that the output of function $\mathscr{G}$ be different. This is exactly the combinatorial balls and bins problem \cite{mitzenmacher2017probability}, in which we seek the number of possible ways to put $n$ indistinguishable balls into $k$ distinguishable bins. Here, bins are the possible values of variables $x_k$s, and balls are the modulation map $\mathscr{E}$. Therefore, using the formula of balls and bins, we get the following upper bound $|\mathcal{R}_s|\leq \binom{K+q-1}{q-1}.$ 
 
 For the lower bound, we should consider a scenario where the outputs of each $\mathscr{E}(x_k)$ overlap with each other. The most overlapping occurs when  $\mathscr{E}$ is a linear function. As a result, for the function $\mathscr{G}$,  we have that 
 \begin{align}
    \mathscr{G}(x_1,\ldots,x_K)=\sum_{k=1}^K\mathscr{E}(x_k) = \mathscr{E} \Big(\sum_{k=1}^Kx_k\Big).
 \end{align}
 Now, because $\mathscr{E}(\cdot)$ is also a bijective mapping,  the cardinality of the range of the function $\mathscr{G}(\cdot)$ equals to the cardinality 
 of the its input domain, i.e. $\sum_{k=1}^Kx_k$. Using that $x_k\in \{0,1,\ldots,q-1\}$, we note that the number of possible values of the range of $\sum_{k=1}^Kx_k$ is $K(q-1)+1$.

\subsection{Proof of Lemma \ref{lem:Lagrangian}}
\label{sec:Prooflem1}
We first rewrite problem \eqref{eq:feasibility-convex} with a new constraint on $\|\bm{s}\|_2^2\leq \tilde{\epsilon}$ as follows: 
\begin{subequations}
\label{eq:feasibility-proof}
\begin{align}
\nonumber
 \tilde{\mathcal{P}} & = {\rm find} ~~~\bm{x}  \\ &~~
{\rm s.t.}~~~
|\Vec{s}_i-\vec{s}_j|^2  \geq |f^{(i)}- f^{(j)}|^2, ~~ \forall (i,j) \in  [M]^2, \\ 
~& ~~~~~~\|\bm{s}\|_2^2 \leq \tilde{\epsilon}.
\end{align}
\end{subequations}
Then, let $\mathcal{L}(\bm{x},\bm{\Lambda},\rho)$ be  the Lagrangian function associated to problem \eqref{eq:feasibility-proof}  as
\begin{align}
    \nonumber \mathcal{L}(\bm{x},\bm{\Lambda},\rho)  = &\langle \mathcal{D}(\bm{Ax}), \bm{\Lambda} \rangle +  \langle \mathcal{D}(\bm{f}), \bm{\Lambda} \rangle \\ \label{eq:lagnrangian} &+ \rho (\bm{x}^{\mathsf{T}}\bm{A}^{\mathsf{T}} \bm{A}\bm{x}-\tilde{\epsilon}),
\end{align}
where $ \mathcal{D} :\mathbb{R}^M \mapsto \mathbb{R}^{M\times M}$   denotes a distance operator, which for vector $\bm{u}\in \mathbb{R}^{M}$ is defined as
\begin{align}
       \mathcal{D}(\bm{u}) := (\bm{u}\odot\bm{u})\mathds{1}_M^{\mathsf{T}} +  \mathds{1}_M(\bm{u}\odot\bm{u})^{\mathsf{T}} - 2\bm{u}\bm{u}^{\mathsf{T}},
\end{align}
in which $\mathds{1}_M$ is a vector of size $M\times 1$ whose all elements are equal to one. Also, the $\bm{\Lambda}$ matrix includes Lagrangian multipliers such that $[\bm{\Lambda}]_{i,j} = \lambda_{i,j}$. Hence, the  Lagrangian cost function $\mathcal{L}(\bm{x}, \bm{\Lambda},\rho)$ in \eqref{eq:lagnrangian} can be equivalent to the following term
\begin{align}
    \nonumber
     \mathcal{L} &= \langle (\bm{Ax}\odot\bm{Ax})\mathds{1}_M^{\mathsf{T}},\bm{\Lambda} \rangle +  \langle  \mathds{1}_M(\bm{Ax}\odot\bm{Ax})^{\mathsf{T}}, \bm{\Lambda} \rangle 
     -  \\ \nonumber
     &~~2 \langle  \bm{A}\bm{x}\bm{x}^{\mathsf{T}}\bm{A}^{\mathsf{T}} \bm{\Lambda} \rangle
     +  \langle \mathcal{D}(\bm{f}), \bm{\Lambda} \rangle \\ & +  \rho (\bm{x}^{\mathsf{T}}\bm{A}^{\mathsf{T}} \bm{A}\bm{x}-\tilde{\epsilon}). 
\end{align}
By reformulating the previous equations, one can reach the following expression.
\begin{align}
\nonumber
    \mathcal{L}(\bm{x}, \bm{\Lambda},\rho) ~&= \langle \bm{x}\bm{x}^{\mathsf{T}},\underbrace{\bm{A}^{\mathsf{T}}(\rho \bm{I} + \bm{\Lambda} - {\rm Diag}(\bm{\Lambda}\mathds{1}_M) )\bm{A}}_{\bm{Z}}\rangle  \\ & + \langle \mathcal{D}(\bm{f}), \bm{\Lambda} \rangle  - \rho \tilde{\epsilon}.
\end{align}
Next, minimizing the above Lagrangian cost function over variable $\bm{x}$ gives us
\begin{align}
   \mathcal{L}(\bm{x}, \bm{\Lambda},\rho) = 
    \begin{cases}
    \label{eq:LagInnerDis}
    \langle \mathcal{D}(\bm{f}), \bm{\Lambda} \rangle  - \rho \tilde{\epsilon},\quad \bm{Z}  \succeq 0,\\
    -\infty, \quad \quad {\rm otherwise},
    \end{cases}
\end{align}
in which the constraint  $ \bm{Z} \succeq 0$  or equivalently  $\bm{Y} := \rho \bm{I} + \bm{\Lambda} - {\rm Diag}(\bm{\Lambda}\mathds{1}_M)\succeq 0$ forces the maximum summation of row of Lagrangian multipliers $\bm{\Lambda}$ to be upper bounded by $\rho$, i.e., $\|\bm{\Lambda}\|_{1\to  1} \leq \rho$. Strictly speaking, the diagonal element $i$ of matrix $\bm{Y}$ is given by
\begin{align}
    \nonumber
    [\bm{Y}]_{ii}  &= [\rho \bm{I} + \bm{\Lambda} - {\rm Diag}(\bm{\Lambda}\mathds{1}_M)]_{i,i} \\ \label{eq:diagelements}& = \rho + \lambda_{i,i} -\sum_{j=1}^M \lambda_{i,j}\,,
\end{align}
where $\lambda_{i,j}$ denotes the $(i,j)$ element of $\bm{\Lambda}$. Since all $\lambda_{i,j}$s are positive numbers, to make the diagonal elements of the matrix $\bm{Y}$  positive, we need to set $\rho \geq \|\bm{\Lambda}\|_{1\to  1}$. Accordingly, we have 
\begin{align}
\nonumber
    \rho  \geq  \|\bm{\Lambda}\|_{1\to  1}  & = \max_{i} \sum_{j=1}^M \lambda_{i,j} \\ & \nonumber \geq \max_{i} \sum_{j=1, j\neq i}^M \lambda_{i,j} \\ \nonumber & \geq \sum_{j=1, j\neq i}^M \lambda_{i,j} \\ \label{eq:rhoInqSum} & =  \sum_{j=1}^M \lambda_{i,j}  -  \lambda_{i,i},
\end{align}
where the last inequality means that all diagonal elements in \eqref{eq:diagelements} are positive. Then, one can write the matrix $\bm{Y}$ as $\bm{Y} = \bm{\Lambda} + \bm{\Gamma}$ where $\bm{\Gamma} = \rho \bm{I} - {\rm Diag}(\bm{\Lambda}\mathds{1}_M)$ is a diagonal matrix with positive elements, i.e., $\bm{\Gamma}_{ii} \geq \lambda_{ii} \geq 0$. On the one hand, matrix $\bm{\Gamma}$ is a diagonal matrix with positive elements, which means it is a PSD matrix. On the other hand, matrix $\bm{\Lambda}$ is a PSD matrix since it is a matrix of the Lagrangian multiplier. Therefore, because  $\bm{\Lambda}$ and   $\bm{\Gamma}$ both are PSD matrices, accordingly,  $\bm{Y} $ is also a PSD matrix.

Next, using this fact yields the following upper bound on the Lagrangian for the PSD case of $\bm{Z}$ where we use H\"{o}lder's inequality on the inner product in \eqref{eq:LagInnerDis} to obtain   
\begin{align}
\nonumber
    \mathcal{L}(\bm{x}, \bm{\Lambda},\rho)  &\leq   \|\bm{\Lambda}\|_{1\to  1}  \sum_{i=1}^M\max_j [\mathcal{D}(\bm{f})]_{i,j}  - \rho \epsilon\\ \nonumber &  \leq  \|\bm{\Lambda}\|_{1\to  1} M \|\mathcal{D}(\bm{f})\|_{\infty}  - \rho \epsilon \\ & \nonumber  \leq  \rho (M \|\mathcal{D}(\bm{f})\|_{\infty} - \epsilon ). 
\end{align}
Hence, to make the problem feasible, the Lagrangian needs to be negative, otherwise maximizing over $\bm{\Lambda}$ leads to  $\infty$ for $\mathcal{L}(\bm{x}, \bm{\Lambda},\rho)$. Therefore, by setting  $\tilde{\epsilon}\geq M \|\mathcal{D}(\bm{f})\|_{\infty}$, or equivalently $\epsilon^{-1}\geq \tfrac{\|\bm{s}\|_2^2}{\|\bm{A}\|} =  \|\mathcal{D}(\bm{f})\|_{\infty},$  the problem is always feasible.

\bibliographystyle{IEEEtran}
\bibliography{IEEEabrv,Ref2}

\end{document}